\documentclass[usenatbib]{elsarticle}
\usepackage[english]{babel}
  \newcommand{\exclude}[1]{}
\usepackage[utf8x]{inputenc}
\usepackage{amsfonts}
\usepackage{amsmath}
\usepackage{amssymb}
\usepackage{xcolor}
\usepackage{enumitem}
\usepackage{graphicx}
\usepackage{gensymb}
\usepackage{fancyhdr}
\usepackage{times}
\usepackage{widetext}
\usepackage{ulem}
\usepackage[caption = false]{subfig}
\usepackage{lineno}

\usepackage{url}
\newcommand{\commentOut}[1]{}

\newcommand{\be}{\begin{equation}}
\newcommand{\ee}{\end{equation}}
\newcommand{\beq}{\begin{eqnarray}}
\newcommand{\eeq}{\end{eqnarray}}

\graphicspath{{./}} 

\begin{document}
\begin{frontmatter}

\title{X-ray annual modulation observed by XMM-\textit{Newton} and Axion Quark Nugget Dark Matter}


\author[2,1]{Shuailiang Ge}
\ead{sge@pku.edu.cn}
\author[1]{Hikari Rachmat}
\author[1]{Md Shahriar Rahim Siddiqui}
\author[1]{Ludovic Van Waerbeke}
\author[1]{Ariel Zhitnitsky}

\address[2]{Center for High Energy Physics, Peking University, Beijing 100871, China}
\address[1]{Department of Physics and Astronomy, University of British Columbia, Vancouver, V6T 1Z1, BC, Canada}

\begin{abstract}
The XMM-\textit{Newton} observatory shows evidence, with a 11 $\sigma$ confidence level, for seasonal variation of the X-ray background in the near-Earth environment in the 2-6 keV energy range \citep{Fraser:2014wja}. The authors argue that the observed seasonal variation suggests a possible link with dark matter. We propose an explanation which involves the Axion Quark Nugget (AQN) dark matter model. In our proposal, AQNs can cross the Earth and emit high energy photons at their exit. We show that the emitted spectrum is consistent with \cite{Fraser:2014wja}, and that our calculation is not sensitive to the specific details of the model. Our proposal predicts a large seasonal variation, on the level of 20-25\%, much larger than conventional dark matter models (1-10\%).
Since the AQN emission spectrum extends up to ∼100 keV, well beyond the keV sensitivity of XMM-\textit{Newton}, we predict the AQN contribution to the hard X-ray and $\gamma$-ray backgrounds in the Earth's environment. The Gamma-Ray Burst Monitor (GBM) instrument, aboard the FERMI telescope, is sensitive to the 8 keV-40 MeV energy band. The  NuSTAR (Nuclear Spectroscopic Telescope Array) is a NASA space based X ray telescope which operates in the range 3 to 79 keV is also sensitive to higher energy bands.  We suggest that the multi-year archival data from the GBM or NuSTAR could be used to search for a seasonal variation in the near-Earth environment up to 100 keV as a future test of the AQN framework.
\end{abstract} 

\begin{keyword}
Axion, Dark Matter,  X-ray 
\end{keyword}

\end{frontmatter}

\section{Introduction}\label{sec:introduction}
The near-Earth Cosmic X-Ray Background (CXB) has been recently measured by \cite{Fraser:2014wja} using archived data from the X-ray XMM-Newton telescope. After a big effort to remove the instrumental background and the constant cosmic background, they discovered a seasonal variation of a strong CXB residual, with a confidence level of 11$\sigma$ in the 2-6 keV energy band  (see Fig.~14 of \cite{Fraser:2014wja}). Moreover, the amplitude of the variation is an order one effect with the signal. If this CXB residual was coming from unresolved extragalactic sources, we would expect it to be uniform and not varying with seasons. The fact that there is an order one effect correlated with the relative positions of the Sun and Earth is unexpected and it suggests a local, near Earth and/or Sun, cause. As a result, the authors of \cite{Fraser:2014wja} searched  for a possible explanation outside of the conventional astrophysical paradigm, after excluding other instrumental possibilities.

The authors   of \cite{Fraser:2014wja}  propose an  explanation based on the assumption that keV axions are emitted by the Sun and convert to X-rays in the Earth's magnetosphere. These X-rays would be subsequently elastically scattered, on average, through a right angle towards the XMM-\textit{Newton} telescope. Note that an even earlier idea was proposed by \cite{DiLella:2003,Davoudiasl:2005nh,Davoudiasl:2008fy} which views the axion-emitting solar core through the solid Earth with an X-ray telescope. The original idea by  \cite{DiLella:2003} cannot explain the effect found by \cite{Fraser:2014wja} because the XMM-\textit{Newton}'s operations exclude pointing at the Sun and at the Earth directly.  
Some of the major criticism of the \cite{Fraser:2014wja} interpretation include the following~\citep{Roncadelli:2014lsa}: a) Due to conservation of momentum, in conventional  cases, the X-ray photons generated in the magnetic field should be collinear with the parent axions. Therefore, since XMM-\textit{Newton} never directly points towards the Sun, it should not see any solar axions; ~
b) Only in the case of a highly inhomogeneous component of the magnetic field with a fluctuation in the keV scale would the photons be non-collinear with the parent axions. Such a fast fluctuating component  is very unlikely to be a dominant portion of a  geomagnetic field. Even if non-collinear effects are generated in the geomagnetic field and we assume that the photon flux converted from axions would be  totally isotropic, the geometric factor $\xi=\Omega_{\rm XMM}/4\pi$ (where $\Omega_{\rm XMM}$ is the aperture of XMM-\textit{Newton}) is very small, $\xi\simeq10^{-5}$. This is in strong disagreement  with the requirement of $\xi\simeq1$ for the interpretation of the observed seasonal variation as  proposed by~\cite{Fraser:2014wja}. Other issues with this interpretation were also discussed in \cite{Roncadelli:2014lsa}.

The main  goal of the present work is to offer an alternative explanation for the observed  seasonal variation \cite{Fraser:2014wja} which does not suffer the fundamental problems mentioned above. Before we discuss our proposal, we would like to make few comments about \cite{Fraser:2014wja} in relation to previous  studies on the CXB.
It is important to point out that the data set used by ~\cite{Fraser:2014wja}  is perfectly consistent with all previous measurements of CXB when averaged over all seasons. The novelty here is the seasonal variation. 
The fact that it is an order of one effect is very unexpected, because it is normally assumed that the isotropic CXB residual is related to unresolved extra-galactic  point sources, therefore uncorrelated with the positions of the Sun or Earth. In this sense the claim by ~\cite{Fraser:2014wja}  on observation of the seasonal variations 
is hard to reconcile  with claim  of ref. \cite{Lumb:2002sw} that  the X ray background measurements in (2-10) keV band could be interpreted that the CHANDRA observations have resolved (70-90)\%  of the background into discrete sources. Our view here  is that there could be some loopholes with this interpretation, the topic which is obviously well beyond the scope of the present work. In what follows, we  assume that the seasonal variations indeed had been observed and properly measured by \cite{Fraser:2014wja}.

\begin{figure}
    \centering
    \includegraphics[width=0.8\linewidth]{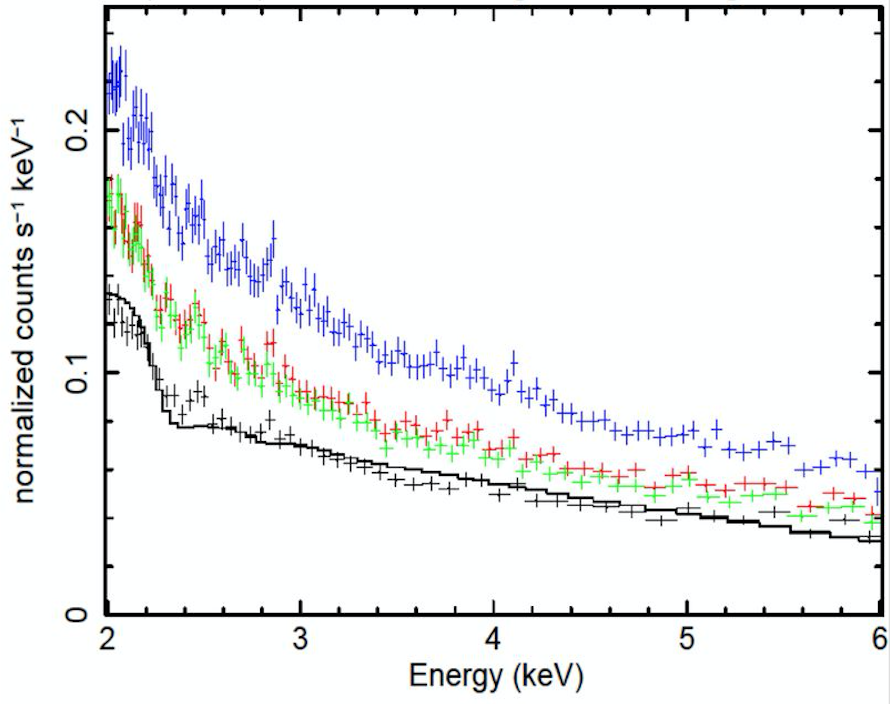}
    \caption{2-6{\rm~keV} X-ray background spectra detected by the EPIC pn camera carried by XMM-\textit{Newton} (the data are integrated from 2000 to 2012) for each of the four spacecraft seasons: Winter (black), Summer (green), Spring (red), and Fall (blue). The solid black line corresponds to the fit (\ref{N_0}) for Winter season. The plot is adopted from Figure 14(a) in~\protect\cite{Fraser:2014wja}.}
    \label{fig:XMMFig14a}
\end{figure}

Although the explanation given by~\cite{Fraser:2014wja} turns out to be untenable, the phenomenon of a seasonally varying X-ray background around the Earth detected with an 11$\sigma$ confidence level remains a mystery (see Fig.~\ref{fig:XMMFig14a}). The seasonal variation pattern is clearly related to the Earth's revolution around the Sun, which strongly indicates that dark matter galactic wind could play a central role. In the present  work we propose that the observed seasonally varying X-ray background could be a result of the annually modulating dark matter wind in the context of the Axion Quark Nugget (AQN) dark matter model \citep{Zhitnitsky:2002qa}.
In our framework, the AQNs emit X-rays isotropically and can propagate in any directions. The radiated X-rays are automatically subject to seasonal variations, since the AQNs are the dominant contributor to dark matter in this framework. Our proposal is therefore very different from \cite{DiLella:2003,Davoudiasl:2005nh,Davoudiasl:2008fy,Fraser:2014wja} which consider axions as the dominant source of dark matter. More importantly, in our proposal the emitted X rays do not scatter through a right angle to reach the telescope, which is the weakest and physically unjustified assumption of the explanation proposed by ~\cite{Fraser:2014wja}.

The AQN model was initially proposed to explain why dark matter and visible matter densities assume similar  magnitudes, $\Omega_{\rm DM}\sim \Omega_{\rm visible}$. The basic features of the model will be reviewed in section \ref{sec:AQN}, but at its heart lies the existence of antimatter nuggets which can interact strongly with regular matter. Remarkably, the antimatter dark matter formulated in the model does not lead to contradictions with known observations. It is quite the contrary; it leads to a series of observational signatures that could explain several outstanding astrophysical puzzles. A non-exhaustive list includes the ``Primordial Lithium Puzzle" \citep{Flambaum:2018ohm}, ``The Solar Corona Mystery"  \citep{Zhitnitsky:2017rop,Raza:2018gpb}, the recent EDGES observations   \citep{Lawson:2018qkc},  and the annual modulation observed by the DAMA/LIBRA experiment\footnote{The DAMA/LIBRA (DL) experiment \citep{Bernabei:2013xsa,Bernabei:2014tqa}
claims an observation of an annual modulation  in the 1-6 keV recoil energy range with a $9.5 \sigma$ confidence level, strongly suggesting that the observed modulation has a dark matter origin. However, the conventional interpretation in terms of WIMP-nucleon  interaction   is excluded by other experiments.    The  AQN framework offers an  alternative source of modulation observed by DL in the form of neutrons which have been liberated from surrounding material~\citep{Zhitnitsky:2019tbh}.}. In the center of the Milky Way, the interaction between antimatter AQNs and baryonic matter also leads to electromagnetic signatures which could explain various emission excesses in different frequency bands, such as the well known 511 keV line~\citep{Oaknin:2004mn,Zhitnitsky:2006tu} and mysterious diffuse    UV radiation \cite{Zhitnitsky:2021wjb}.


The basic idea in our proposal follows from the fact that antimatter AQNs hit the Earth at a low rate. These AQNs will lose some of their mass from annihilation, and they will also lose some momentum ~\citep{Lawson:2019cvy}. The nuggets are not completely destroyed. At the moment of their exit, they are very hot objects as a result of friction and annihilation events occurring in the Earth's deep underground layers. At the exit point, their temperature can be as high as $T_0\sim$ 200 keV (this is discussed in Section \ref{sec:model_built}). The AQNs slowly cool down while they continue their trajectory away from the Earth's surface and emit radiation. At this stage, the AQNs continue to lose their accumulated heat and slowly decrease their internal  temperature. 
On average, when AQNs reach distances of the order $r\gtrsim 8 R_{\oplus}$, their  temperature remains very high,  $T\gg 10~ {\rm keV}$. The XMM-\textit{Newton} operates precisely at such distance and can easily observe these X-rays emitted by AQNs. AQNs represent the dominant form of dark matter in this model, and the velocities of AQNs hitting the Earth are different for different seasons. We call this the annual modulation of the dark matter wind (see footnote \ref{season} for a comment on this terminology). As a result, the flux  of AQNs that leave the Earth also depends on seasons, which consequently  leads to a seasonal varying X-ray background. 
Ideally, our model should match the observed seasonal variation shown in Fig.~\ref{fig:XMMFig14a}. This can only be done with the knowledge of the telescope orientation and position changes as a function of seasons, information we do not have. For this reason, a precise comparison with observations cannot be done at this time\footnote{\label{season}There are subtle points here related to XMM-\textit{Newton}'s  position and its  view angle as it orbits the Earth. This complication does not allow an immediate interpretation of the data in terms of the conventional annual modulation, which is normally attributed to dark matter wind with its maximum on June 1 and minimum on Dec 1. See the original paper by \cite{Freese:1987wu} and  review \cite{Freese:2013} for more information. The maximum and minimum values will obviously get shifted as a function of the satellite's position with respect to the Earth's surface. We will make a few comments on these complications later in the text.}, and future observations are required to calculate a precise X ray seasonal variation consistent with the telescope orientations and positions changes over time. However, we will show that our model can reproduce the amplitude of the seasonal variations, which is on the level of 20-25\% (see Fig.~\ref{fig:XMMFig14a}), and the energy spectrum measured in \cite{Fraser:2014wja}.

The paper is organized as follows. In Section~\ref{sec:AQN}, we briefly review the AQN model. In Section~\ref{sec:model_built}, we calculate the thermal emission spectrum of a single AQN and its cooling process in space after it leaves the Earth. In Section~\ref{sec:theoretical_predictions}, we calculate the total spectrum received by XMM-\textit{Newton} from AQNs that enter the field of view of the telescope, and compare it with the observations. In Section~\ref{sec:seasonal_variation}, we calculate the seasonal variation of the AQN-induced spectrum. Finally, we conclude the main points of this work in Section~\ref{sec:conclusion}. 

\section{The AQN model}\label{sec:AQN}
In this model, dark matter consists of axion quark nuggets (AQNs) which are macroscopically large objects of nuclear density with a typical size of $R\sim10^{-5}$~cm and typical mass of $B\sim 10^{25}$ times the mass of a proton. $B$ is also called the baryonic charge. AQNs are dense objects made of standard model quarks and gluons in the colour superconducting (CS) phase \citep{Zhitnitsky:2002qa}.   This model  is conceptually similar to the original nuclearites proposed by \cite{Witten:1984rs}, where these nuggets are ``cosmologically dark" not through the weakness of their 
interactions but due to their small cross-section to mass ratio. 
As a result, the corresponding  constraints on this type of dark matter place a lower bound on their mass, rather than their coupling constant. 

\exclude{There are several additional elements in the AQN model that we can compare with the older and well-studied nuclearite by \cite{Witten:1984rs}:
 
\begin{itemize}
    \item The AQNs are formed from the collapse of closed axion domain walls which are copiously produced during the QCD transition in the early Universe. An AQN is squeezed by an axion domain wall (with its QCD substructure) as its shell, which acts adds an additional stabilization factor. This helps to alleviate a number of 
the problems inherent to  the original Witten proposal\footnote{\label{first-order}In particular, in the original proposal by \cite{Witten:1984rs}, the first order phase transition was the required feature of the construction of the nuclearite.  However, it is known that the QCD transition is a crossover rather than a first order phase transition. It should be contrasted with the AQN framework
because the first order phase transition is not required, as the axion domain wall plays the role of the squeezer.  Furthermore, it has been argued that Witten's nuclearites 
are likely to evaporate on the Hubble time-scale even if they were formed. In the AQN framework, 
a fast evaporation does not occur because the AQNs are stable as a result of additional external pressure from the axion domain walls. In contrast, Witten's nuclearites are stable objects at zero pressure.}.  
    \item The nuggets can   be made of matter as well as  \textit{antimatter} (to be coined as the   \textit{antinuggets}). The direct consequence of this feature is that   the visible and dark matter abundances must   assume the  same order of magnitude  without any fine tuning, i.e. $\Omega_{\rm visible}\sim\Omega_{\rm dark}$. This relation is a natural result of the AQN framework, which is not sensitive  to any specific parameters such as axion mass or nugget size, as explained below. 
\end{itemize}
}

The presence of a large amount of antimatter in the form of high density AQNs leads to  many  
 observable consequences   as a result of   very rare annihilation events between antiquarks in AQNs and  baryons in the visible Universe. 
    
It is normally assumed that the Universe started in a neutral phase with zero baryonic charge, then it evolved into a state with a net positive baryon number through a ``baryogenesis'' process. In the AQN model, the ``baryogenesis" is replaced by the \textit{charge separation process} in which the total baryon charge of the Universe remains zero at all times. However, due to the global  $\cal CP$ violating processes associated with the axion's potential misalignment angle $\theta_0\neq 0$ during 
the early formation stage at the QCD scale, the number and size distributions of nuggets and antinuggets will necessarily be different by an order of one. This happens regardless of the axion mass $m_a$ and the initial value of $\theta_0$.
 We refer the readers to some original work for a detailed analysis  on the formation of nuggets~\citep{Liang:2016tqc}, the development of nugget-antinugget asymmetry \citep{Ge:2017ttc,Ge:2017idw}, and their size distribution and survival pattern \citep{Ge:2019voa}  in the unfriendly environment of the early Universe, see also a brief recent review \cite{Zhitnitsky:2021iwg} covering these topics. 
 \exclude{
 The asymmetry in numbers and size distributions between nuggets and antinuggets leads naturally to a Universe where dark matter is composed of nuggets and antinuggets (with respective cosmological mass densities $\Omega_{N}$ and $\Omega_{\bar N}$). There is a preference for the antinuggets by an order of one, while the remaining matter constitutes the observed regular baryonic matter with mass density $\Omega_{\rm visible}$ (stars, galaxies, gas, etc.). The resulting baryon charge $B$ of the Universe remains zero, and no specific baryogenesis mechanism is necessary while we have $\Omega_{\rm visible}\sim \Omega_{\rm dark}$ and
 \begin{eqnarray}
 \label{Omega}
 \Omega_{\rm visible}&=&\Omega_{\bar{N}}-\Omega_{N}\nonumber\\
 \Omega_{\rm dark}&=&\Omega_{\bar{N}}+\Omega_{N}.
 \end{eqnarray}
Eq. (\ref{Omega}) is very generic and a robust consequence of the AQN framework: where both components  $\Omega_{\rm visible}$ and $\Omega_{\rm dark}$  originate from the same  QCD physics at the same cosmological epoch, and both are proportional to the same fundamental scale, the $\Lambda_{\rm QCD}$. This fact provides a natural mechanism, which is very different from the Weakly Interacting Massive Particle's ``miracle," by which why dark matter and baryons mass abundances are comparable.

\subsection{Astrophysical signatures of the model}
Unlike conventional dark matter candidates such as Weakly Interacting Massive Particles (WIMPs), the presence of antimatter in the antinuggets makes them strongly interacting with baryonic matter. Intuition dictates that such a model would be in strong contradiction with existing astrophysical observations. However, detailed studies of the AQN's interaction in astrophysical environments show that the model does not contradict any known observational constraints on dark matter or antimatter and that the very small number density of AQNs prevents the emission mechanisms to be overwhelming. The main reasons for AQNs to behave as Cold Dark Matter despite their strong, but extremely rare, interaction with baryons can be summarized as follows \citep{Zhitnitsky:2006vt}:

\begin{itemize}
    \item The typical baryon charge carried by a nugget is huge. $|B|\sim 10^{25}$ (constrained by observations and will be discussed later in this section), which implies that the number density of nuggets is extremely low.
    \item 
    The key ratio relevant for cosmology $\sigma/M\sim 10^{-10}{\rm~cm}^2/{\rm g}$  is far below the astrophysical and cosmological limits $\sigma/M < 1{\rm~cm}^2/{\rm g}$. Therefore, the AQN qualifies as a Cold Dark Matter candidate.
    
    \item The nuggets are stable objects over the cosmological timescale. AQNs survive in the unfriendly environment of the early Universe, before and after the Big Bang Nucleosynthesis (BBN) epoch \citep{Ge:2019voa}. A dominant portion of the AQNs also survive violent events such as galaxy formation and star formation.
    \item The nuggets have a very large binding energy, so the quarks/antiquarks locked in the core cannot participate in BBN, which happens at $T\sim 0.1$~MeV. Therefore, the conventional BBN picture holds   with possible small corrections on the order of $10^{-10}$, which in fact could resolve the primordial lithium puzzle~\citep{Flambaum:2018ohm}.
    \item Due to the small ratio $\sigma/M\sim B^{-1/3}\ll 1$, the nuggets completely decouple from photons, so the conventional picture of structure formation holds.
    \item The nuggets do not modify the conventional analysis of the CMB. They provide possible small radiation corrections which could resolve the tension between standard predictions and the EDGES observation (stronger than anticipated 21 cm absorption features)~\citep{Lawson:2018qkc}.
\end{itemize}
}

In the AQN model, dark matter will emit radiation when it collides with visible  baryons as a result of annihilation processes. Because of the AQN's low number density, this is a very rare event. 
\exclude{When it happens, only the surface of the AQN will radiate, while most of the matter inside the nugget remains hidden and dark. For this reason, it is expected that AQNs with a larger baryon charge, $\left<B\right>$, will generate even weaker radiation\footnote{To be more specific, only antinuggets play an important role here in observations, since they carry a huge amount of antimatter that could annihilate with matter from the visible Universe. Also, it is antinuggets that are relevant for our present study of the varying X-ray background. Therefore, the nuggets or AQNs discussed later in this paper refers to antinuggets, if not specified.}. It is also expected that baryon rich environments, such as the early Universe, the core of a galaxy, planets, and stars will emit more AQN related radiation for a given baryon charge, $\left<B\right>$, than in a baryon poor environment.
}
It is surprising at first that a model with dark matter in form of the antimatter nuggets  could even pass the simplest observational constraint, but the reason lies in the fact that most of a nugget's mass is inside the nugget and does not contribute to the emission processes. A comparison of the emission mechanisms to astrophysical observations from radio to $\gamma$-ray wavelengths suggests that the nugget's mass should have a baryon charge of $\left<B\right> > 10^{24}$ to avoid the overproduction of the observed galactic diffuse background and be consistent with all known observations. This corresponds to an AQN mass of only $\sim 1$ gram, therefore AQNs of mass $\sim 1$ gram and higher are viable dark matter candidates. As mentioned in the Introduction, there are several excesses of emissions in different frequency bands contained in the galactic spectrum, which seem to be consistent with the moderate emission processes inherent to the AQN model. 
\exclude{The best known example is the strong galactic 511 keV line. Several of these diffuse emissions could be explained within the framework of the AQN model if the nuggets carry a baryon charge of order $\left<B\right> \sim 10^{25}$. We refer the readers to~\cite{Oaknin:2004mn,Zhitnitsky:2006tu,Forbes:2006ba}
 for further details with  explicit computations in different frequency bands. In all of these cases, the emitted photons are generated in the outer layer of the nuggets, consisting of electrons (positrons in case of antinuggets), known as the electrosphere. 
}

The X-ray emission in the near-Earth environment, which is the subject of our present work, is   originated from the electrosphere of the nugget. Therefore the thermal properties of the electrosphere plays a crucial role. The relevant thermal features of the electrosphere have been analyzed previously in  \cite{Forbes:2008uf} in the context of galactic emission, where the nugget's internal  temperature turns out to be very low, being around $T\sim$~eV.
due to the very low density of the environment on the level $n_B\sim 10^2 {\rm cm^{-3}}$. 
In our present study, we are interested in the  nuggets crossing   the Earth's interior with a very  high density of the surrounding material, around $n_B\sim 10^{24} {\rm cm^{-3}}$, and even higher in the Earth's core. As a result,   the nuggets crossing the Earth's interior will acquire very  high temperatures, reaching up to $T\simeq$ 200  keV, as estimated in the next section.   
For such  high temperatures,  several new phenomena related to ionization, plasma frequency,  and other many-body effects, which had been previously neglected in   ~\cite{Forbes:2008uf}, become very important and have to be explicitly incorporated into the computational framework. The corresponding modifications of the dynamics of the electrosphere accounting for all of these effects  will be the subject  of  the following Section~\ref{sec:model_built}.
We use these results in Section~\ref{sec:theoretical_predictions} to calculate the spectrum accumulated by XMM-\textit{Newton} from the hot AQNs based on the observatory's configuration and orbit information, and compare it with the observations.

\section{AQN-induced x-rays}\label{sec:model_built}
In order to theoretically calculate the spectrum received by XMM-\textit{Newton} from the radiation of hot AQNs that have crossed  the Earth's interior, the first step is to know the radiation spectrum from the  electrosphere of  an AQN characterized by a high temperature $T\simeq$ 200 keV, which represents the topic of   subsection~\ref{subsec:radiation}. In subsection~\ref{subsec:cooling}, we examine the cooling process of AQNs in space after they leave the Earth. Since the AQN's radiation features change as its temperature drops, we need to know the temperature, intensity, and spectrum  of AQNs when they reach the region $r\sim 10 R_{\oplus}$,  where the  XMM-\textit{Newton} is operational. 

\subsection{AQN emissivity}\label{subsec:radiation}
The properties of thermal emission from the electrosphere of a nugget have been discussed in~\cite{Forbes:2008uf}.
The process of radiation is Bremsstrahlung emission when the positrons scatter and emit photons with typical energy $\sim T$.   First, we will briefly summarize the previous results here. Next, we will discuss a number of complications which are relevant for our present work (when the temperature is very high $T\simeq$ 200-500 keV). These were ignored in previous studies with  $T\simeq $ eV in the context of galactic emission.

The spectral surface emissivity is denoted as $dF/d\omega=dE/dtdAd\omega$, representing the energy emitted by a single nugget per unit time, per unit area of the nugget's surface, and per unit frequency. It has the following expression~\citep{Forbes:2008uf}:
\begin{equation}
\label{eq:dFdomega0}
    \frac{dF}{d\omega} =\frac{1}{2}\int_{0}^{\infty}dz~\frac{dQ}{d\omega}(\omega,z), ~~~~  F_{\text{tot}}(T) \equiv \int_0^{\infty} \frac{dF}{d\omega} d\omega \simeq 
  \frac{16}{3}
  \frac{T^4\alpha^{5/2}}{\pi}\sqrt[4]{\frac{T}{m}},
\end{equation}
where
\begin{equation}\label{eq:dQ0}
    \frac{dQ}{d\omega}=n^2(z)\cdot \frac{4\alpha}{15}\left(\frac{\alpha}{m_{e}}\right)^2 2\sqrt{\frac{2T}{m_{e}\pi}} \left(1+\frac{\omega}{T}\right){\rm e}^{-\omega/T} h\left(\frac{\omega}{T}\right).
\end{equation}
$n(z)$ is the local density of positrons at distance $z$ from the nugget's surface, which has the following expression
\begin{equation}\label{eq:nz0}
n(z)=\frac{T}{2\pi\alpha}\frac{1}{(z+\bar{z})^2}, 
\end{equation}
with 
\begin{equation}\label{eq:zbar}
    \bar{z}^{-1}=\sqrt{2\pi\alpha}\cdot m_{e} \cdot \left(\frac{T}{m_{e}}\right)^{1/4}, ~~~~ n(z=0)\simeq \left(m_eT\right)^{3/2},
\end{equation}
where $n(z=0)$ reproduces an approximate  formula  for the plasma density in the Boltzmann regime at temperature, $T$. The  parameter $\bar{z}$   was introduced in ~\cite{Forbes:2008uf} as a constant of integration and has the dimension of a length scale where the Boltzmann regime is realized.
The function $h(x)$ in Eq. (\ref{eq:dQ0}) is a  dimensionless  function computed in~\cite{Forbes:2008uf} (see ~\ref{appendix:F_details} for details). The important features of the spectrum will be discussed in detail at the end of this subsection, but we would like to emphasize that it is qualitatively different from conventional black body radiation, despite of the fact  that the electrosphere is characterized by a specific  temperature, $T$. 
The reason is that the size of the system is much smaller than the photon's mean free path and, as a result, the photons cannot thermalize in this system. The resulting spectrum is very broad and extends to the energies much lower than the temperature $T$, in contrast with black body radiation.

A typical internal temperature $T$ of  the nuggets can be estimated from the condition that
 the radiative output of equation (\ref{eq:dFdomega0}) must balanced the flux of energy onto the 
nugget  due to the annihilation events. In this case we may write, 
\be
\label{eq:rad_balance}
F_{\text{tot}}(T) \cdot (4\pi R^2)
\simeq \bar{\kappa}\cdot  (\pi R^2) \cdot (2~ {\rm GeV})\cdot n_{\rm out}  \cdot v_{\rm AQN},  
\ee 
where the left hand side accounts for the total energy radiation from the nuggets' surface per unit time as given by (\ref{eq:dFdomega0}), while    the right hand side  accounts for the rate of annihilation events when each successful annihilation event of a single baryon charge produces $\sim 2m_pc^2\simeq 2~{\rm GeV}$ energy. In  (\ref{eq:rad_balance}) we assume that  the nugget is characterized by the geometrical cross section $\pi R^2$ when it propagates 
in environment with local density $n_{\rm out} $ with velocity $v_{\rm AQN}\sim 10^{-3}c$.

The factor $\bar{\kappa}$ is introduced to account for the fact that not all matter striking the nugget will 
annihilate and not all of the energy released by an annihilation will be thermalized in the nuggets.
 In a neutral dilute environment considered  previously \cite{Forbes:2008uf}  the value of $\bar{\kappa}$ cannot exceed $\bar{\kappa} \lesssim 1$ which would 
correspond to the total annihilation of all impacting matter into to thermal photons. The high probability 
of reflection at the sharp quark matter surface lowers the value of $\bar{\kappa}$. The propagation of an ionized (negatively charged) nugget in a  highly ionized plasma (such as solar corona)   will 
 increase  the effective cross section. As a consequence,   the value of  $\bar{\kappa}$ could be very large as discussed in \cite{Raza:2018gpb} in application to the solar corona heating problem.

The thermal properties presented above were applied to the study of the emission from AQNs from the Galactic Centre, where a nugget's internal temperature is very low, $T\sim$ eV, as already mentioned  in Section~\ref{sec:AQN}. 
When the nuggets propagate in the Earth's atmosphere, the AQN's internal temperature starts to rise up to $\sim 20$ keV or so. When the AQN enters the Earth's surface, it further heats up to $\sim$ 200 keV, due to the much higher density of the Earth's interior. Indeed, from (\ref{eq:dFdomega0}) and (\ref{eq:rad_balance}) one can estimate the internal temperatures in atmosphere and in the Earth's interior as follows:
 \be
\begin{aligned}
 \label{T}
 T_{\rm atmosphere}&\sim& 15 ~{\rm keV} \cdot \left(\frac{n_{\rm atmosphere}}{3\cdot 10^{20} ~{\rm cm^{-3}}}\right)^{\frac{4}{17}}\left(\frac{ \bar{\kappa}}{0.1}\right)^{\frac{4}{17}},   \\
  T_{\rm interior}&\sim& 200 ~{\rm keV} \cdot \left(\frac{n_{\rm interior}}{10^{25} ~{\rm cm^{-3}}}\right)^{\frac{4}{17}}\left(\frac{\bar{\kappa}}{0.1}\right)^{\frac{4}{17}},
  \end{aligned}
 \ee 
where  the temperature $T\sim 15$ keV had been used previously in \citep{Zhitnitsky:2020shd} for the AQN propagating under thunderclouds to explain the unusual Cosmic Ray like events  observed by Telescope Array Collaboration, while  $T\sim 200$ keV had been used in \cite{Zhitnitsky:2021qhj} to explain unusual clustering events observed by the HORIZON -10T collaboration. In what follows, we use $T_0\approx 200 $ keV as the benchmark temperature for the nuggets exiting the Earth's surface. However, as we explain in next subsection  a precise  computations  of $T_0$ from the first principles is a very difficult problem. Therefore, we opted to keep $T_0$ as the phenomenological parameter, which will be determined by observational constraints.

  While the  nuggets propagate underground, the heat from the electrosphere is transferred to the nugget's core, accompanied by the emission of photons with a spectrum characterized by Eq. (\ref{eq:dFdomega0}). These processes are very complicated to compute because in this temperature regime, a number of many-body  effects in the electrosphere, that were previously ignored, become important. In what follows, we explain the physics of these effects, while all of the technical details are developed in  \ref{appendix:F_details}. 
 
1. \textit{The modification of the positron's  density $n(z)$ in the electrosphere}\\ 
The most important modification due to high temperature occurs as a result of the ionization of the system. Loosely bound positrons leave the system, and strongly bound positrons change their positions and momenta to adjust to the corresponding modifications of the system. Indeed, 
 the neutrality of the AQN will be lost due to the ionization at $T\neq 0$,  in which case  the   antimatter nuggets will acquire a negative electric charge due to the ionized positrons. 
 \exclude{The corresponding charge, $Q$, for AQNs can be estimated as  follows \citep{Zhitnitsky:2017rop,Ge:2019voa}:
  \be
  \label{Q1}
Q\simeq 4\pi R^2 \int^{\infty}_{z_1}  n(z)dz\sim \frac{4\pi R^2}{2\pi\alpha}\cdot \left(T\sqrt{2 m_e T}\right), 
  \ee
 where $n(z)$ is the density of the positrons (\ref{eq:nz0}) in the electrosphere.   In this estimate, it is  assumed that the  weakly  bound positrons, with binding energy $\epsilon\lesssim T$, will be stripped off of the electrosphere as a result of high temperature, $T$. These  loosely bound positrons 
 are localized mostly at the outer region of the electrosphere, at distances 
 $z> z_1(T)\approx (2 m_e T)^{-1/2}$, which motivates the cutoff in our estimate (\ref{Q1}).
}

Since the temperature of the AQN's core becomes very high due to the large number of annihilation events in the Earth's interior, a large number of weakly bound positrons will be stripped off of the nugget, and the number density of remaining positrons will drastically decrease. 
The corresponding changes in the electrosphere are determined by nontrivial non-equilibrium dynamics, which shall not be discussed in the present work. Instead, we introduce a phenomenological suppression factor, $\kappa \ll 1$, not to be confused with parameter $\bar{\kappa}$ introduced earlier, 
which effectively accounts for the relevant physics:  
\begin{equation}\label{eq:nz}
    n(z)=\kappa\cdot \frac{T}{2\pi\alpha}\frac{1}{(z+\bar{z})^2}.
\end{equation}
Although the precise computations of the coefficient  $\kappa \ll 1$ from first principles
is not possible at this stage, a simple estimate based on simple assumptions will be given below. The estimate suggest that $\kappa \sim 10^{-3}$  as a result of dramatic expansion of the electrosphere at high temperatures, in which case the effective density of the positrons (\ref{eq:nz}) will be dramatically diminish.

Eq. (\ref{eq:nz}) is a simplification which does not take into account the fact that loosely bound positrons will be completely stripped off by high temperature, while more strongly bound positrons will be less affected and stay bound. One can easily add this feature to our simplified analysis by describing $\kappa$ as a step function: 
\begin{equation}
\label{kappa}
\kappa(z,T)=
\begin{cases}
0 ~~\text{if }~~ z\geq z_1~   \\
\kappa(T)~~ \text{if} ~~ z < z_1
 \end{cases},
\end{equation}
  where $z_{1}(T)$ is defined as 
\begin{equation}\label{eq:z_1}
    z_{1}(T)\simeq\frac{1}{\sqrt{2m_{e}T}}.
\end{equation}
 
\begin{figure}
    \centering
    \includegraphics[width=1\linewidth]{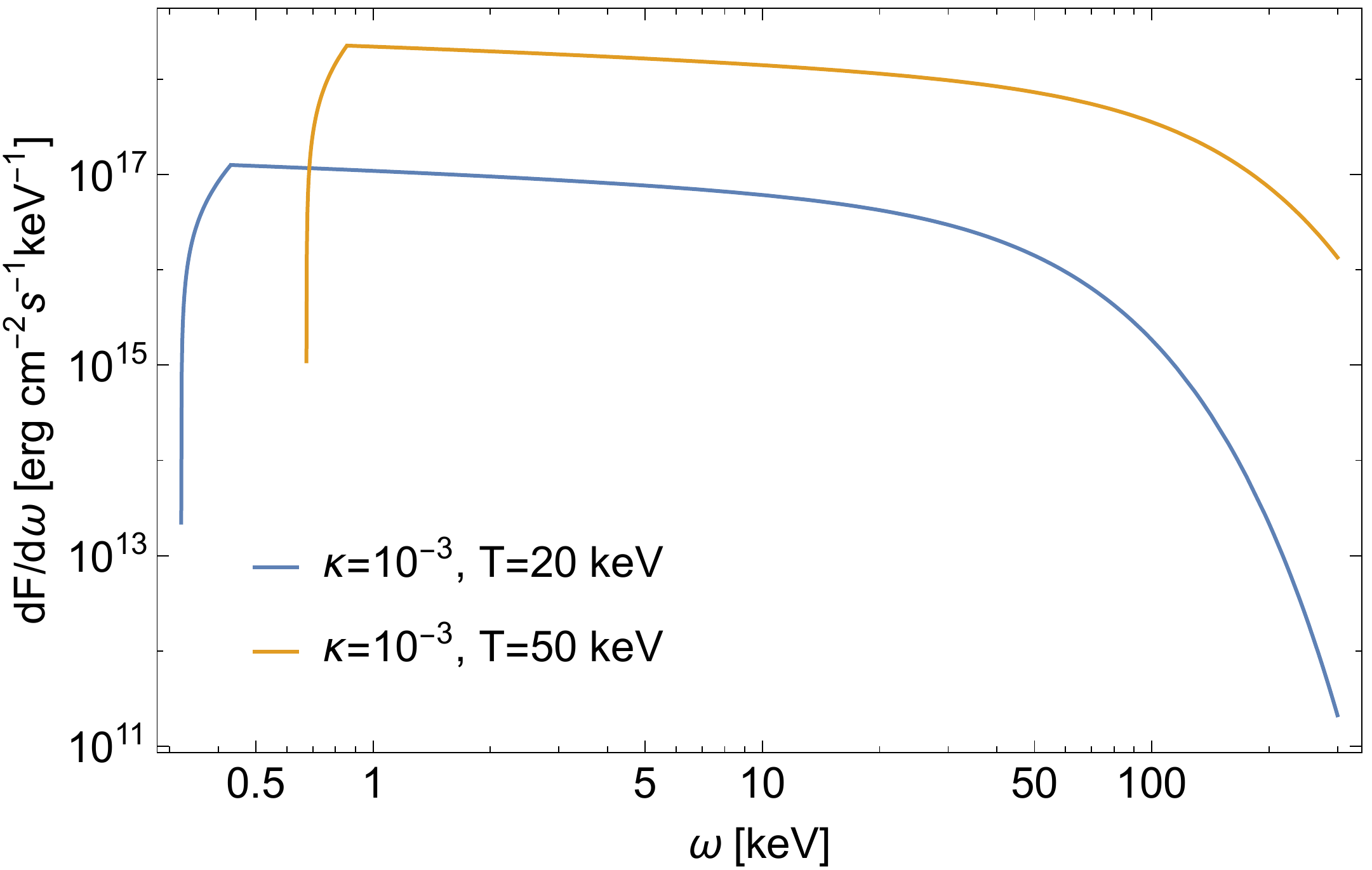}
    \caption{The spectral surface emissivity of a nugget with all of the effects discussed in this subsection included, see (\ref{eq:dFdomegaexpression}) in Appendix~\ref{appendix:F_details}. 
     This is plotted for $\kappa = 10^{-3}$ with $T=20$~keV and $50$~keV respectively, for illustrative purpose.
    }
    \label{fig:dFdomega}
\end{figure}
In this way, we preserve the crucial feature of the system that loosely bound positrons from the outer region of the electrosphere are stripped from the nugget and do not  participate in the cooling of the system.

2. \textit{The role of the plasma frequency} \\
The plasma frequency $\omega_{p}$ characterizes the propagation of photons in a plasma. It can be thought of as an effective mass for the photons: only photons with an energy larger than this mass can propagate outside of the system, while  photons with $\omega<\omega_{p}$ can only propagate for a short time and distance $\sim\omega_{p}^{-1}$ before being absorbed back. For our estimates, we will use a conventional non-relativistic expression  for $\omega_{p}$:
\begin{equation}\label{eq:omega}
    \omega_p^2(z)=\frac{4\pi\alpha n(z)}{m_{e}}; ~~~ \omega_p (z)\simeq  \sqrt{\frac{2T}{m_{e}}}\frac{\sqrt{\kappa}}{(z+\bar{z})}, 
\end{equation}
where we substituted Eq. (\ref{eq:nz}). 

The important implication of the plasma frequency $\omega_p (z)$ 
is that the densest regions of the electrosphere stop emitting photons 
because the plasma frequency is too high, since $\omega^2_p\sim n$ according to (\ref{eq:omega}). This implies that the emissivity (Eq. (\ref{eq:dQ0})) from the dense regions will be exponentially suppressed at the plasma frequency \citep{Forbes:2008uf}:
\exclude{
One more comment is that when calculating $dQ/d\omega$~(\ref{eq:dQ0}), we should not use the full density $n(z)$~(\ref{eq:nz}). This is because the number density that participate in generating the free-propagating photons is constrained by the factor $\sim\theta(\tilde{\epsilon}_{p_{1}}+\tilde{\epsilon}_{p_{2}}-\omega)\cdot \theta(\omega-\omega_{p})$ [$\theta(x)$ is the unit step function], see the Appendix A4 in~\cite{Forbes:2008uf} for more details. The factor $\theta(\tilde{\epsilon}_{p_{1}}+\tilde{\epsilon}_{p_{2}}-\omega)$ accounts for energy conservation: the initial energy must be larger than the energy of emitted photon; and the factor $\theta(\omega-\omega_{p})$ accounts for the effect of plasma frequency: only photons with $\omega>\omega_{p}$ can propagate freely. It turns out that the number density $\tilde{n}(z)$ that participate in generating the free-propagating photons is suppressed exponentially compared with the full density $n(z)$~(\ref{eq:nz}), which gives $\tilde{n}^{2}(z)\sim \exp[-\omega_{p}(z)/T]\cdot n^{2}(z)$~\citep{Forbes:2008uf}. So now $dQ/d\omega$ becomes }
\begin{equation}\label{eq:dQdomega}
    \frac{d\tilde{Q}}{d\omega}(\omega,z)\sim {\rm e}^{-\omega_{p}(z)/T}\cdot  \frac{dQ}{d\omega}(\omega,z).
\end{equation}

\bigskip

With all of these effects taken into account, the spectral surface emissivity (\ref{eq:dFdomega0}) can now be  numerically computed. We refer the reader to~\ref{appendix:F_details} for technical details of the computations. 
Two examples are shown in Fig.~\ref{fig:dFdomega} with $\kappa (T)= 10^{-3}, T=20$~keV and  $\kappa (T)= 10^{-3}, T=50$~keV respectively, for illustrative purpose.
Fig.~\ref{fig:dFdomega} reveals some important features. First of all,  the spectrum is almost flat in the region $\omega\lesssim T$. This is a direct manifestation of a very generic property of emission by charged particles when the energy of the emitted photon is much smaller than all of the other scales of the problem, the so-called ``soft Bremsstrahlung" emission or ``soft photon theorem." In this case, the emission is known to show a $d\omega/\omega$ behaviour for the probability to emit a soft photon with frequency $\omega$. This property implies that the intensity of radiation,  $dF/d\omega\sim {\rm constant}$, must be flat for soft photons. As we will discuss in Section \ref{sec:theoretical_predictions}, this unique property of the spectrum  will play a key role in our interpretation of the spectrum observed by XMM-\textit{Newton}. On the other hand, for large $\omega\gg T$, the exponential suppression, $\exp(-\omega/T)$, becomes the most important feature of the spectrum. The complete suppression of the emission 
   at very small $\omega\ll T$ is an artifact of our simplification of the density, $n(z)\sim\kappa(z)$, in form of  a step function (\ref{kappa}).  There is another   cusp  behaviour  also at $\omega\ll T$ (peak on Fig. ~\ref{fig:dFdomega}). This results from our simplified treatment of the plasma frequency, $\omega_p$, when the $dF/d\omega$ is approximated by a piecewise function (when the  emission with $\omega \geq \omega_p$ from a high density region   occurs with no suppression, while emission with $\omega \leq \omega_p$ from the same region is completely  dropped). In reality, both of these effects leading to the cusps should be described by a smooth function.  However,  this part of the spectrum with $\omega\ll T$ will not play any role in our analysis which follows\footnote{An  important consequence of the strong suppression at small $\omega\ll T$ is that the intensity of the visible light emission with $\omega\sim 1$ eV is  strongly suppressed in comparison to the X-ray emission. It could play a dramatic role in the identification of AQN annihilation events in the atmosphere with the so-called skyquakes. They occur when a sonic boom is not accompanied by any visible light, which would  normally be expected for any meteors-like events, see \cite{Budker:2020mqk} for details.}.
  
  The next step is the computation of the cooling rate, done in subsection \ref{subsec:cooling}. For this purpose,
  we need the total surface emissivity, $F_{\rm tot}(T, \kappa)$, as a function of $T$ and $\kappa$. 
This is done by numerically  integrating $dF/d\omega$ over $\omega$.  The difference in comparison with simple analytical formula  given in (\ref{eq:dFdomega0}) is that
the expression below accounts for all additional  effects  discussed in this subsection.     The technical details of the calculations can be found in  \ref{appendix:F_details}, Eq. (\ref{eq:Ftotfittingappendix}).
We parameterize  the final formula for the emissivity, which will be used in subsection \ref{subsec:cooling} to study the AQN's cooling, as follows:
\begin{equation}\label{eq:Ftotfitting}
    F_{\rm tot}(T)\simeq\frac{\alpha}{15\pi^{5/2}}\frac{T^{5}}{m_{e}} \cdot c_{1}(\kappa)\left(\frac{T}{10{\rm~keV}}\right)^{c_{2}(\kappa)}
\end{equation}
with
\begin{equation}\label{eq:c12}
    c_{1}(\kappa)=4\kappa^2,~~~~ c_{2}(\kappa)=-0.89.
\end{equation}

\subsection{AQN cooling}\label{subsec:cooling}
While passing through the Earth, nuggets will heat up by friction and annihilation events. Their temperature when exiting the Earth's surface is denoted by $T_0$, and it assumes the value $T_0\sim 200$ keV as estimated in (\ref{T}). While traversing the Earth, the nugget will heat up in a fraction of a second because of the very efficient energy transfer between the nugget and its surrounding dense material. However, it is expected that $T_0$ cannot become  higher than $\sim$ 500 keV because of different processes. These include $e^+e^-$ pair production and black body radiation, which start to dominate the nugget's emission deep underground, and become much more important than the Bremsstrahlung radiation (\ref{eq:dQ0}).  Calculating $T_0$ precisely from first principles remains very difficult because the energy transfer in the Earth's interior includes complicated processes, such as  turbulence and acoustic shock waves
with a very large Mach number, $M$. $M=v_{\rm AQN}/c_s\gg 10$, where $c_s$ is the speed of sound of the surrounding material. This part of the nugget's physics is very complicated, and it is not part of our present work. For this reason, we will treat $T_0$ as a phenomenological parameter, which we expect to be close to $T_0\simeq 200 $ keV,
as mentioned above.
However, for illustrative purposes we also present the  results for $T_0\simeq 500$ keV. 

Fortunately, these  complications do not affect our analysis once the AQNs exit the surface and start to travel in empty space. After exiting Earth, the energy loss from the AQN into space is entirely determined by Eq. (\ref{eq:Ftotfitting}). In this case, the initial condition for the cooling is simply characterized by $T_0$. One can completely ignore any new annihilation events at this point because the density of the material in Earth's atmosphere drops very quickly with height. Therefore, the nuggets are assumed  to be travelling  in empty space immediately after they exit the Earth's surface, with initial temperature $T(r=R_{\oplus})=T_0$.  

Our goal now is to calculate the energy loss rate of the heated AQN while it travels through space, away from Earth, with a typical dark matter speed of $\sim 220 {\rm km/s}$. The total initial energy accumulated by the AQN is determined by its exit temperature, $T_0$, and specific heat, $c_V$. 
\exclude{The corresponding expression for unpaired quark matter is known \citep{Alford:2007xm} and it is given by:
\be
\label{quark-matter}
c_V=\frac{N_cN_f}{3}\mu^2 T,
\ee
where $\mu$ is the chemical potential, and $N_c, N_f$ are the number of colours and flavours in the system.
}There are many different CS phases with drastically different expressions for $c_V$. In particular, in 2SC (two flavour superconducting phase), the expression for the specific heat  assumes the form \citep{Alford:2007xm}:
\be
\label{2SC}
c_V\simeq   \frac{1}{3}T (\mu_d^2+\mu_u^2),
\ee
where chemical potentials in CS phases are in the range $\mu_u\simeq \mu_d\simeq 500$ MeV. This numerical value  is perfectly   consistent with our studies of the typical  value of  the AQN's chemical potential  at the moment of its formation \citep{Ge:2019voa}. 
For our numerical analysis in what follows, we use expression (\ref{2SC}).

The energy of the nugget decreases when its temperature decreases, according to the conventional formula  
\begin{equation}\label{eq:cooling}
    dE=c_{V}\cdot V \cdot dT,
\end{equation}
where   $V$ is the AQN's   volume. 
The energy emitted by a nugget per unit time has been computed in the previous section and it is given by  (\ref{eq:Ftotfitting}): 
\be 
\label{eq:cooling1}
-dE/dt=F_{\rm tot}(T)\cdot 4\pi R^2,
\ee
where sign  minus  implies that the energy of the AQN is decreasing with time as a result of emission. Combining (\ref{eq:cooling}) and (\ref{eq:cooling1}), we arrive at
the desired equation  describing the change  of the temperature, $T$, with time, $t$,
while the AQN is moving away from the Earth and emitting photons:
\begin{equation} \label{eq:dTdt}
    \frac{dT}{dt}=-\frac{4\pi R^2}{V}\frac{1}{c_{V}(T)}F_{\rm tot}(T).
\end{equation}
The solution of this differential equation, with initial condition $T(t=0)=T_0$, is given by:
\begin{equation}\label{eq:T_t}
\begin{aligned}
&\left(\frac{t}{1{\rm~sec}}\right)\simeq 
\frac{0.34}{c_{1}(\kappa)[c_{2}(\kappa)+3]}
\left(\frac{R }{10^{-5}{\rm~cm}}\right) \left(\frac{\mu_{u,d}}{500{\rm~MeV}}\right)^2\\
&~~~~~\cdot \left[\left(\frac{T}{10{\rm~keV}}\right)^{-[c_{2}(\kappa)+3]}
-\left(\frac{T_0}{10{\rm~keV}}\right)^{-[c_{2}(\kappa)+3]}\right]  ,
\end{aligned}
\end{equation}
where $T(t)=T_0$ at $t=0$, when the nugget exits the Earth's surface. We refer the readers 
to Appendix~\ref{appendix:cooling} for the  details  on the derivation. 

\begin{figure}
    \centering
    \includegraphics[width=1\linewidth]{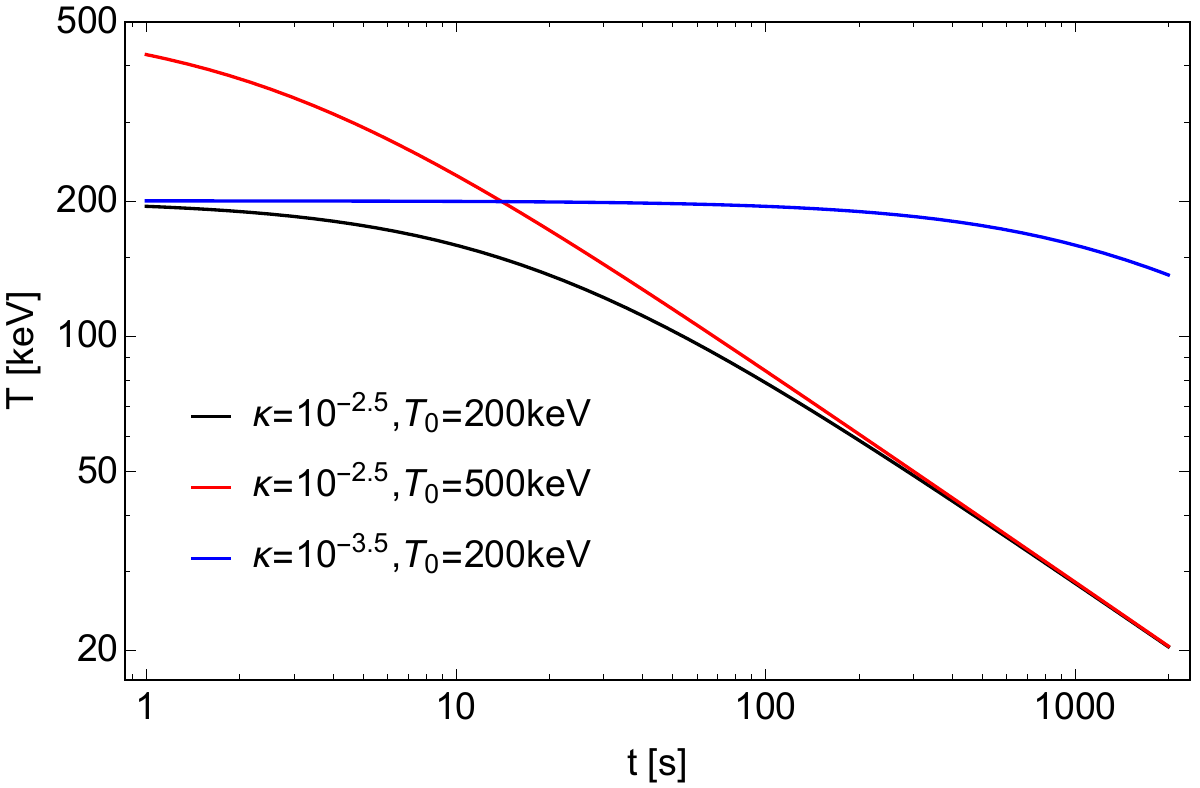}
    \caption{$T$ vs. $t$ for different values of $\kappa$ and $T_0$. $T=T_0$ at $t=0$. An important feature here is that the behaviour $T(t)$ at $t\geq 100$ s (when XMM-\textit{Newton} becomes operational) is not sensitive to the initial value of $T_0$, see text for explanations.}
    \label{fig:Tt}
\end{figure}

Fig.~\ref{fig:Tt} shows $T$ as a function of time, $t$, for different values of $\kappa$ and $T_{0}$. We choose $R =2.25\cdot 10^{-5}$~cm, which has been previously used in axion emission studies \citep{Lawson:2019cvy}. Fig.~\ref{fig:Tt} illustrates a very important result: after $t\approx 100$ seconds, when the AQN is at distance $r\geq 3 R_{\oplus}$, the temperature $T(t)$ is not very sensitive to the initial temperature  $T_0$ for a given coefficient $\kappa$. This is because AQNs with higher initial temperature $T_0$ emit more radiation and cool down more quickly. As a result, $T(t)$ is much more sensitive to $\kappa$ than $T_0$, as shown by the blue and black lines in Fig.~\ref{fig:Tt}. This is because a smaller value of $\kappa$ leads 
 to a drastic reduction of the emission.  As a consequence of this suppressed emission, the temperature remains close to its initial value, $T_0$, for a  long period of time, $t\sim 10^3$ seconds. \cite{Fraser:2014wja} selected observations such that XMM-\textit{Newton} would always point away from the Earth and Sun. Therefore, we expect that their signal will be weakly sensitive to $T_0$ and strongly sensitive to $\kappa$.
 
 \exclude{
Another important quantity is the AQN energy loss rate. It can be computed as follows: the total energy stored in a nugget at the moment of exit is
\begin{equation}
\label{E_0}
    E_{0}=\int_{0}^{T_{0}} c_{V}(T)VdT.
\end{equation}
The energy lost due to thermal emission to space is given by
\begin{equation}
    E(t)=4\pi R^2  \int_{0}^{t} dt F_{\rm tot}[T(t)],
\end{equation}
where $F_{\rm tot}[T]$ is determined by Eq. (\ref{eq:Ftotfitting}) and $T(t)$ by Eq. (\ref{eq:T_t}). 
The stored  energy  $[1-E(t)/E_{0}]$ as function of time for different values of  $\kappa$  and $T_{0}$ is shown in Fig.~\ref{fig:Et}. This  function describes the fraction of energy remaining in the AQN's core at time $t$, which vanishes when $t\rightarrow \infty$. Fig.~\ref{fig:Et} shows that a smaller $\kappa$ corresponds to a reduced emission and therefore a much slower energy loss rate. For instance, for $\kappa=10^{-3.5}$, the stored energy in the AQN's  core is almost unaltered up to $t\simeq 10^2$ seconds.

\begin{figure}
    \centering
    \includegraphics[width=1\linewidth]{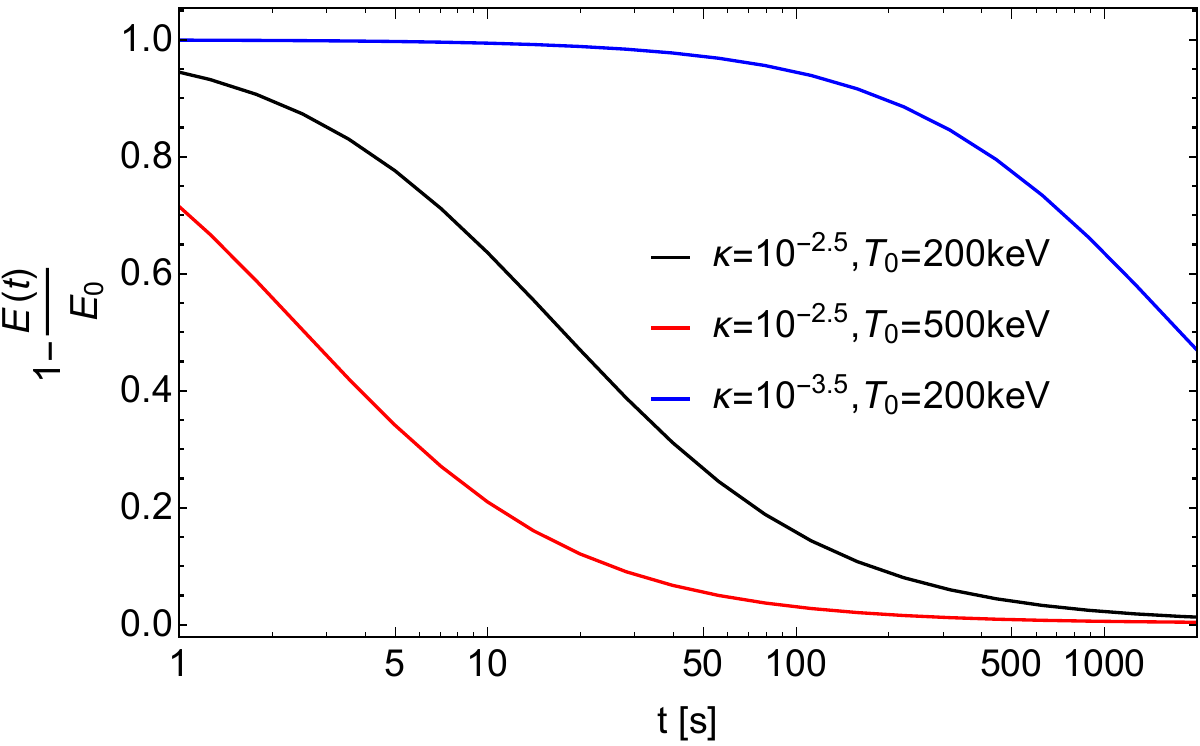}
    \caption{The relative stored energy $\left[1-E(t)/E_0\right]$ vs. $t$, for different values of $\kappa$ and $T_0$. An important feature here is that a smaller  $\kappa=10^{-3.5}$ corresponding to a reduced emission leads to a much slower decay rate. In this case,  the  AQN  keeps its initial energy value up to  $t\lesssim 10^2$~ s.}
    \label{fig:Et}
\end{figure}
}
We conclude this section with a few comments on   the   parameters $T_0$ and $\kappa$, which appear in the  computations and provide a benchmark for our numerical estimates. As we shall see in the next section, the spectrum and the intensity of the emission   depend on these  parameters in a very nontrivial way.  
As already mentioned at the beginning of this section, the exit temperature is expected to be in the range $T_0\sim$ 200  keV according to (\ref{T}). 
\exclude{This results from the very high rate of annihilation events in the dense environment \footnote{Indeed, according to Eq. (\ref{E_0}), the energy, $E_0\simeq \frac{1}{2}c_VVT_0^2$ with $T_0\simeq 500$ keV, is achieved when the AQN travels a distance of order $L\sim 0.5$ km, at which the accumulated annihilation energy, ${(2~\rm GeV})n_B \pi R^2 L$ with $n_B\sim 10^{24} {\rm cm^{-3}}$, becomes the same  order of magnitude as $E_0$.}, and that the heat loss from $e^+e^-$ pair production and black body radiation prevents $T_0$ from going beyond $\sim$ 500 keV.}
The other parameter which enters our computations is the suppression factor, $\kappa$, defined by Eq.  (\ref{eq:nz}). This was introduced to account for the drastic decrease of the positron number density from the electrosphere, which can emit photons. This strong suppression is a direct consequence of high internal temperature, $T_0$, when a large number of  weakly bound positrons     are expanded over much larger distances order of $R$ rather  than distributed over  much shorter distances  of order $\bar{z}$ defined by eq.(\ref{eq:zbar}). The corresponding estimates had been carried out in  Appendix A in \cite{Zhitnitsky:2021qhj} which    explicitly shows the emergence of very small factor  $ (mR)^{-1/2}$ as a manifestation of  this effect, i.e.
\be
\label{kappa-estimate}
\kappa\sim \left(\frac{\bar{z}}{R}\right)^{\frac{1}{2}}\sim \frac{1}{\sqrt{mR}}\frac{1}{\sqrt[4]{2\pi\alpha}} \sim 2\cdot 10^{-3}. 
\ee
Again,  we opted to treat  $\kappa$ as   a free phenomenological parameter in the rest of the paper, similar to the  parameter $T_0$ discussed above.  The  benchmark value for $\kappa$ is given by  (\ref{kappa-estimate}).

\section{Computation of the spectrum and Comparison with XMM-\textit{Newton} data}\label{sec:theoretical_predictions}
This is the central section of the present work, as we 
  are in a position  to compute  the spectrum and intensity received by XMM-\textit{Newton} from the thermal emission of nuggets computed in the previous section. 

We start with the simplified assumption that the nuggets are \textit{uniformly}  distributed around the Earth. We will also assume that the nuggets exit the Earth \textit{radially}.
As we shall see below, we are able to reproduce the spectrum observed by XMM-\textit{Newton} with the AQN framework. Since the calculations are relatively insensitive to the free parameters of the model, our result represents a very generic consequence of the system when the spectrum is essentially determined by the ``soft photon theorem," as we already mentioned at the end of subsection \ref{subsec:radiation}.

\begin{figure}
    \centering
    \includegraphics[width=1\linewidth]{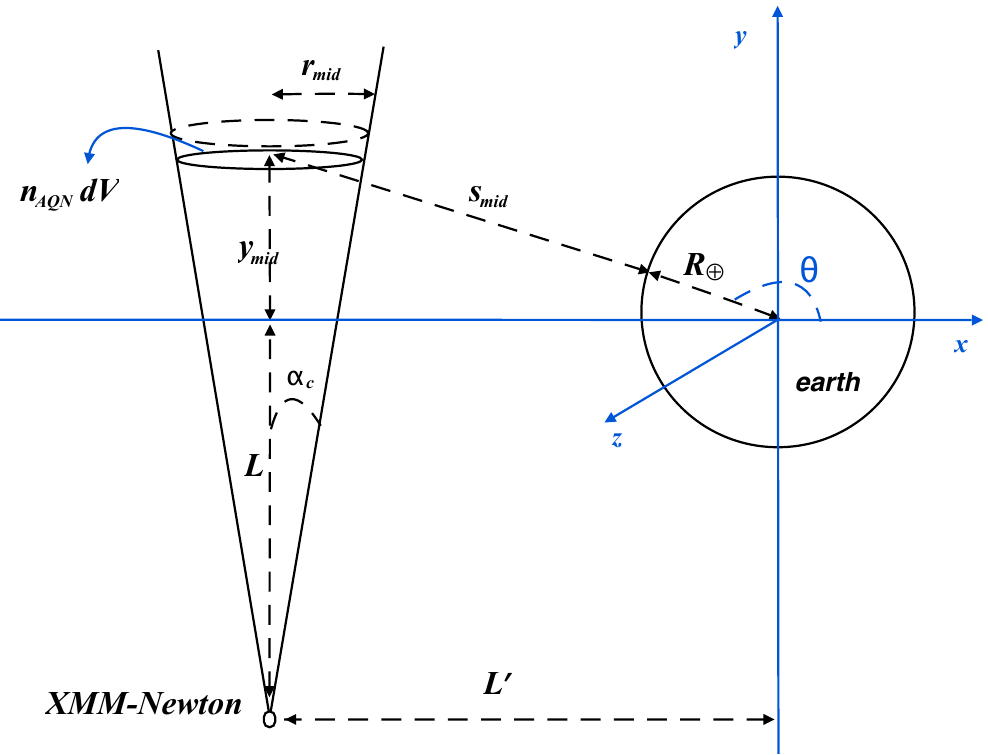}
    \caption{The XMM-\textit{Newton} observatory is assumed to be located at the position $(x,y,z)=(-L',-L,0)$. The cone is the field of view of the EPIC pn camera carried by XMM-\textit{Newton}. In our present work, we focus on this camera (see footnote~\ref{foot:XMMcameras} for details). The cone points in the direction $+y$. $dV$ is the volume of the thin disk, and the number of nuggets contained inside is $n_{\rm AQN}dV$. $r_{\rm mid}$ is the radius of the thin disk. Since the opening angle of the cone is very small, $\alpha_{c}=0.25\deg$, it is a good approximation that all nuggets inside $dV$ are located at the same point $(-L', y_{\rm mid}, 0)$. $y_{\rm mid}$ could be negative, so they have the same distance, $s_{\rm mid}$, to the Earth's surface. The range of $\theta$ is apparently $[\pi/2, \pi+\arctan{(L/L')}]$.
    }
    \label{fig:xmmsketch}
\end{figure}

In what follows, we compute the spectrum and the intensity with the assumptions formulated above.  The corresponding result on the spectrum is not very sensitive to the specific geometry since the X-ray emission from the nugget is isotropic and does not depend on the direction of velocity, nor the orientation with respect to Earth. At the same time, the intensity is highly sensitive to the parameters as we discuss below.  With these simplifications  in mind,   the number density of nuggets that have passed through the Earth is
\begin{equation}\label{eq:nAQN}
    n_{\rm AQN}(s)=\frac{1}{4\pi(R_{\oplus}+s)^2}\cdot \frac{\mathcal{F}}{v_{\rm out}},
\end{equation}
where Fig. \ref{fig:xmmsketch} shows the geometry of the configuration. In Eq. (\ref{eq:nAQN}), $s$ denotes the distance from the Earth's surface. $v_{\rm out}$ is the nugget's velocity leaving the Earth's surface, which is assumed to be the same for all nuggets and independent of $s$. For simplicity, we approximate $v_{\rm out}\simeq v_{\rm in}$, the nugget's velocity when it hits the Earth, although the nuggets may be slowed down by the interactions with their surroundings inside the Earth. This approximation is good enough for our analytical treatment in this section. The effect of the velocity difference between $v_{\rm out}$ and $v_{\rm in}$ will be discussed in the next section where this difference plays a key role in our  studies of  the seasonal variations. We should also note that $n_{\rm AQN}$ is a season invariant, see \ref{appendix:dFdw} with details.  We denote $\mathcal{F}$ as the total nugget flux (number per unit time) that hits the Earth. It has been estimated as follows  ~\citep{Lawson:2019cvy}:
\begin{equation}
\label{F}
    \mathcal{F}\simeq 0.67~{\rm s}^{-1} \left(\frac{\rho_{\rm DM}}{0.3~{\rm GeV}/{\rm cm}^3}\right) \left(\frac{v_{\rm in}}{220{\rm~km/s}}\right)\left(\frac{10^{25}}{\left<B\right>}\right).
\end{equation}
We adopt the following  values for our numerical estimates: $v_{\rm out}\simeq v_{\rm in}\simeq 220$ km/s; average baryon charge, $\left<B\right>=10^{25}$ (which corresponds to an average size of the nugget, $\left<R\right>=2.25\cdot 10^{-5}$~cm). This corresponds to a total flux of $\mathcal{F}\simeq 0.67 {\rm s^{-1}}$ ~\citep{Lawson:2019cvy}.

\begin{figure}
    \centering
    \includegraphics[width=1\linewidth]{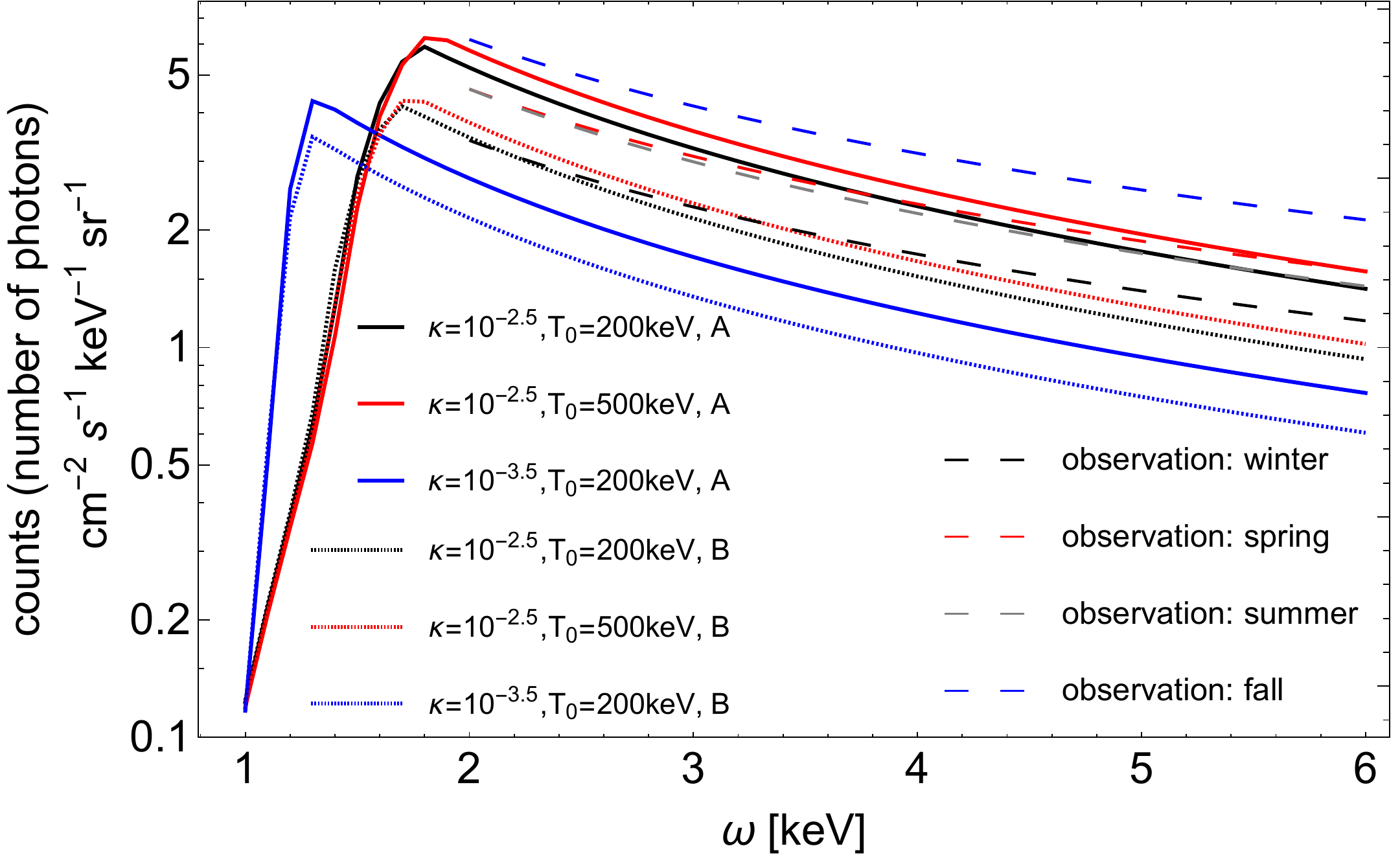}
    \caption{The relation $f$ vs. $\omega$, (\ref{eq:f_xmm}), for $(\kappa,T_0)=(10^{-2.5}, 200{\rm~keV})$, $(10^{-2.5}, 500{\rm~keV})$, and $(10^{-3.5}, 200{\rm~keV})$ respectively. The x-axis represents frequency. The y-axis represents the values of $f^{({\rm theory})}$, given by (\ref{eq:f_xmm}), which is the number of photons received by the XMM-\textit{Newton} observatory (camera EPIC pn) per unit time, per unit area, per unit frequency, and per unit solid angle in the AQN framework. The solid lines are for the case that the XMM-\textit{Newton} observatory locates at position A: $(x,y,z)=(-7,-7,0)R_{\oplus}$. 
    In comparison, we also plotted the spectra for the second option that the observatory locates at position B:  $(x,y,z)=(-8,-6,0)R_{\oplus}$ with the cone orientation keeping unchanged, for the same three groups of $(\kappa,T_0)$, which are denoted as dotted lines. We also plot $f^{({\rm obs})}$, given by  (\ref{eq:fobservation}), representing the data observed by the EPIC pn camera for four seasons, shown  by the four dashed lines respectively.}
    \label{fig:counts}
\end{figure}

\begin{figure}
    \centering
    \includegraphics[width=1\linewidth]{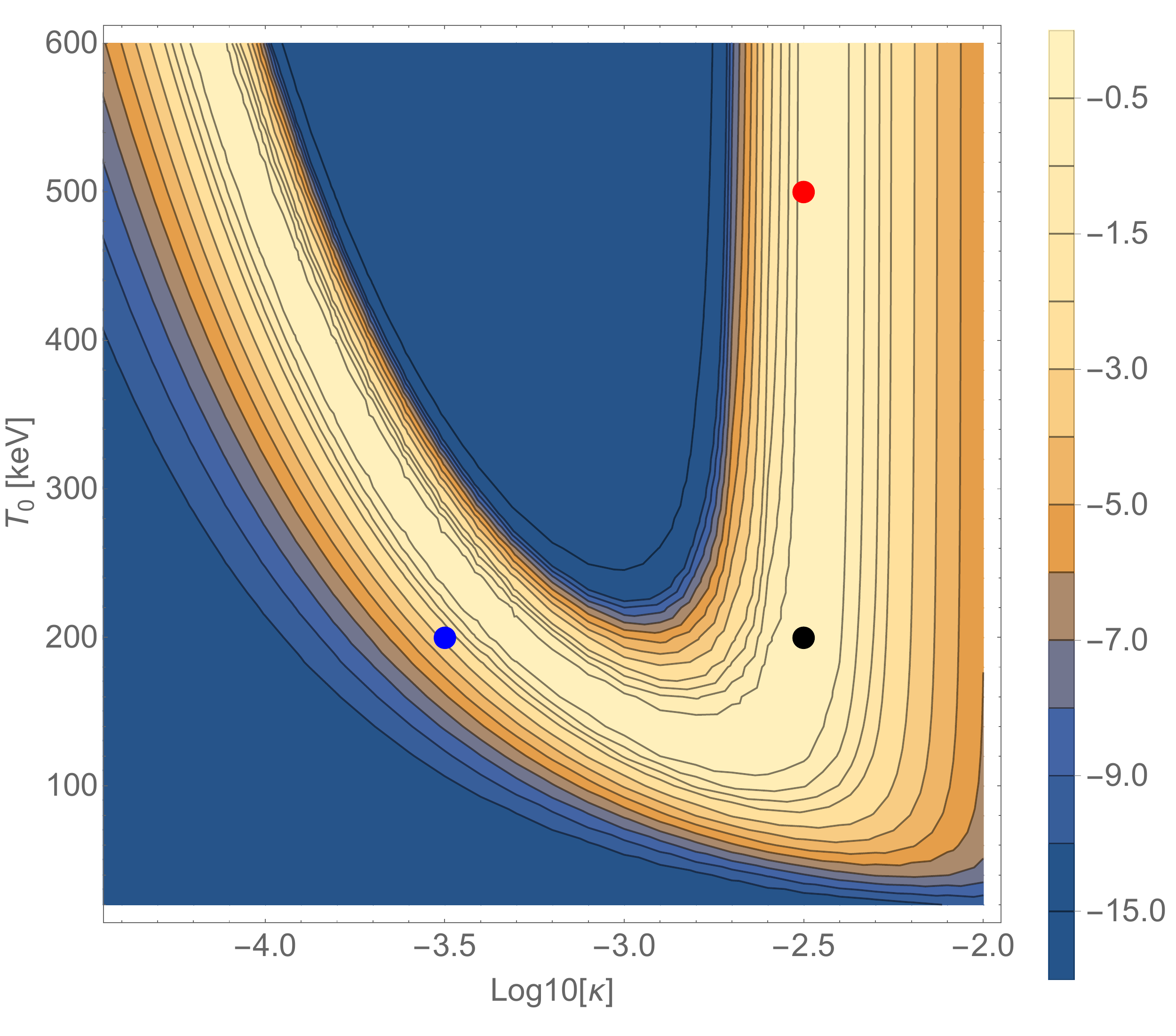}
    \caption{ The contour plot of $2\ln\mathcal{L}(\kappa, T_0)$. The numbers on the color bar represent the values of $2\ln\mathcal{L}$. The three points marked on the plot are the three sets of $(\kappa,T_0)$ that we have chosen in all of our previous plots: $(10^{-2.5}, 200{\rm~keV})$, $(10^{-2.5}, 500{\rm~keV})$, and $(10^{-3.5}, 200{\rm~keV})$.}
    \label{fig:contour}
\end{figure}

Fig.~\ref{fig:xmmsketch} shows the positions of the Earth and the XMM-\textit{Newton} observatory.
It also shows how the XMM-\textit{Newton} observatory receives the radiation from a large number of nuggets, with number density $n_{\rm AQN}$, surrounding the Earth. Various configuration parameters are defined in Fig.~\ref{fig:xmmsketch}'s caption. The following geometric relations are useful:
\begin{equation}
\begin{aligned}
    &s_{\rm mid}(\theta)=\frac{L'}{-\cos\theta}-R_{\oplus},~~y_{\rm mid}(\theta)=-L'\tan\theta, \\
    &r_{\rm mid}(\theta)=(y_{\rm mid}+L)\tan\alpha_{c}.
\end{aligned}
\end{equation}
The number of nuggets inside the thin disk of the cone (shown in Fig.~\ref{fig:xmmsketch}) is
\begin{equation}
    dN(\theta)=n_{\rm AQN}dV=n_{\rm AQN}[s_{\rm mid}(\theta)]\cdot \pi r_{\rm mid}^2(\theta) dy_{\rm mid}(\theta).  
\end{equation}
The distance, $s_{\rm mid}$, and the time, $t$, are connected by the nugget velocity, $v_{\rm out}$: $s_{\rm mid}=v_{\rm out}t$. The spectrum received by the XMM-\textit{Newton} observatory can be calculated as:
\begin{equation}
\begin{aligned}\label{eq:dFrdomega}
    \frac{dF_{r}}{d\omega}&=\int_{\rm cone} \frac{dF}{d\omega}
    \left(T_{\rm mid}, \omega \right)\frac{R^2}{[y_{m}(\theta)+L]^2} ~dN(\theta)\\
    &=\int_{\frac{\pi}{2}}^{\pi+\arctan{(L/L')}}d\theta~
    \left\{\frac{dF}{d\omega}\left(T_{\rm mid}, \omega \right)\frac{R^2}{[y_{m}(\theta)+L]^2}  \right.\\
    &\left.
    \cdot  n_{\rm AQN}[s_{\rm mid}(\theta)]\cdot \pi r_{\rm mid}^2(\theta) \frac{L'}{\cos^2\theta}\right\},
\end{aligned}
\end{equation}
where $\frac{dF}{d\omega}\left(T_{\rm mid}, \omega \right)$ is the spectral surface emissivity 
computed at the moment $t_{\rm mid}=\left[\frac{s_{\rm mid}(\theta)}{v_{\rm out}}\right]$ with the corresponding temperature, $T_{\rm mid}$. The   computations of $\frac{dF}{d\omega}(T, \omega) $ for  arbitrary $T$ have been carried out in 
  section \ref{subsec:radiation}, see Fig. \ref{fig:dFdomega} 
  for example. 
  The $R$   in Eq. (\ref{eq:dFrdomega}) is the nugget's radius, $R\simeq 2.25\cdot 10^{-5} {\rm cm}$, corresponding to $\left<B\right>=10^{25}$. For numerical estimates, we consider two different options (please refer to Fig.~\ref{fig:xmmsketch} for the geometry): first we consider the observatory locating at position $A$: $(x,y,z)=(-7,-7,0)R_{\oplus}$, where $R_{\oplus}$ is the Earth's radius. 
  Secondly we consider the observatory locating at position B:  $(x,y,z)=(-8,-6,0)R_{\oplus}$ with the cone orientation keeping unchanged, for the same three groups of $(\kappa,T_0)$. We observe very minor changes as a result of this modification of the geometry\footnote{The orbit of XMM-\textit{Newton} is highly elliptical, with an apogee altitude of $\sim115000$ km and a perigee altitude of $\sim 6000$ km. The orbit period is $\sim48$ hr. The orbit changes with time, due to several perturbations. We refer the readers to the XMM-\textit{Newton} Users Handbook~\citep{XMMhandbook} for details. The observatory only works at altitudes above the Earth's radiation belts $\sim46000$ km, see, e.g.,~\cite{XMMhandbook,santos2001life}. Therefore, in this section, we choose $L=7R_{\oplus}$, which implies that the altitude of the observatory is $\sqrt{2}L-R_{\oplus}\approx57000$ km.}, see Fig. \ref{fig:counts}. These modifications are much smaller than changes due to  variation of  parameters $T_0$ and $\kappa$ which cannot be computed precisely, as explained in the previous section. Therefore, later when we make the contour plot (Fig.~\ref{fig:contour}) to explore the parameter space of $\kappa$ and $T_0$ and when we discuss the seasonal variation, we consider only single option A as any small modifications of location  lead to very minor variations for the intensity. One should emphasize that the spectrum of the emission is not sensitive to any changes of the parameters as we explain below.
  
The radiation spectrum $dF_{r}/d\omega$ given by (\ref{eq:dFrdomega}) is the energy received by the observatory per unit time, per unit area, and per unit frequency. In order to make a precise comparison  between  our calculations and  the observations \citep{Fraser:2014wja}, we convert $dF_{r}/d\omega$ to $f^{(\rm theory)}$, the \textit{number of photons} received by the observatory per unit time, per unit area, per unit frequency, and \textit{per unit solid angle}, which is defined as follows: 
\begin{equation}\label{eq:f_xmm}
    f^{(\rm theory)}\equiv \frac{1}{\Omega_{c}}\frac{1}{\omega}\frac{dF_{r}}{d\omega},
\end{equation}
where $\Omega_{c}=2\pi(1-\cos\alpha_{c})\approx 5.98\times10^{-5} {\rm~sr}\approx 0.196~\deg^2$ is the solid angle of the cone. The corresponding theoretical 
prediction is plotted in Fig.~\ref{fig:counts},   for several typical values of the parameters of the system,   $\kappa, T_{0}$,  as discussed  in  Section~\ref{sec:model_built}.

In order to compare with observations from XMM-\textit{Newton} in the 2-6 keV energy band, we use the power-law fit,  as given by~\cite{Fraser:2014wja}, see Eq. (11) from that paper\footnote{The symbols in Eq.(11) of ~\cite{Fraser:2014wja} conflict with ours, so we rewrite Eq. (11) as~(\ref{eq:fobservation}), using our own symbols to avoid confusion.}:
\begin{equation}\label{eq:fobservation}
    f^{({\rm obs})}=N_{0}\left(\frac{\omega}{{\rm keV}}\right)^{-\Gamma} \frac{1}{\rm cm^2\cdot  s\cdot keV \cdot sr}.
\end{equation}
The normalization factor  $N_{0}$ is dimensionless, while $f$ is measured in [cm$^{-2}$s$^{-1}$keV$^{-1}$sr$^{-1}$]. For the EPIC pn camera carried by XMM-\textit{Newton}
\footnote{\label{foot:XMMcameras}XMM-\textit{Newton} carries three cameras that are relevant to us: EPIC pn, EPIC MOS1, and EPIC MOS2. The three cameras all clearly show the seasonal variance of the X-ray background with similar values of $(N_{0},\Gamma)$ listed in the main text, which can be seen in Table~3 of~\cite{Fraser:2014wja}. Therefore, we only need to focus on one camera, which is enough for our purpose in the present work to compare the AQN-based calculations with the observations. We choose the EPIC pn camera because it has the largest photon grasp (effective area $\times$ aperture), which is a key parameter in studying the background of X-ray radiation, and because it has better counting statistics than the two EPIC MOS cameras~\citep{Fraser:2014wja}.},
 the values of the numerical  parameters $(N_{0}, \Gamma)$ are: 
\begin{equation}
\begin{aligned}\label{N_0}
{\rm Winter} (N_{0}, \Gamma)&=&(6.66,0.97);\\
{\rm Spring} (N_{0}, \Gamma)&=&(9.08,0.98);\\
{\rm Summer} (N_{0}, \Gamma)&=&(9.60, 1.06);\\
{\rm Fall} (N_{0}, \Gamma)&=&(12.09,0.97),
\end{aligned}
\end{equation}
see Table 3 in~\cite{Fraser:2014wja}. These numbers are obtained by fitting the data observed by the EPIC pn camera (from Fig.~$\ref{fig:XMMFig14a}$) showing the seasonal variation of the X-ray background with 11$\sigma$ significance. The maximum amplitude of the  seasonal variation from these data occurs between Winter and Fall, rather than between Winter and Summer. It has been discussed in ~\cite{Fraser:2014wja} and will be discussed in the context of the AQN model in section \ref{subsec2:seasonal_variation}.

Fig.~\ref{fig:counts} shows our theoretical prediction (the solid lines and dotted lines from Eq. (\ref{eq:f_xmm})) against the observed spectra (the dashed lines from Eq.~(\ref{eq:fobservation})) for four seasons. 
 The similarity between the observations and theoretical computations is impressive, considering that the shape of the predicted radiation spectrum is only slightly sensitive to the parameters $\kappa$ and $T_0$. This result is a direct consequence of the AQN framework\footnote{\label{foot:cusp}A cusp behavior in the region $\omega$ = 1-2 keV in Fig.~\ref{fig:counts} has no physical significance. Rather, it is a reflection of our simplified treatment of the regions with small $\omega$, which results in such a cusp singularity, see comments on this cusp behaviour  in subsection \ref{subsec:radiation}.}.
 The basic reason for the robustness of our prediction is that
 the spectrum shape is essentially determined by the very fundamental ``soft photon theorem," with a specific behaviour, $d\omega/\omega$, for $\omega\ll T$, as we 
 already  emphasized earlier at the end of subsection \ref{subsec:radiation}. The slope, $\Gamma$, as seen in Eq. (\ref{eq:fobservation}), is indeed very close to $\Gamma\simeq 1$ for all seasons. This shows very strong support for our AQN framework, as the spectrum is very robust consequence of the framework, not sensitive to specific details of the computations. 
 
 The intensity of the emission, on the other hand, is sensitive to the parameters  $(\kappa, T_0)$. It is also sensitive to the dark matter distribution, nugget size distribution, velocity distribution, etc., as one can see from Eq. (\ref{F}) for the AQN flux. The distance and orientation of the XMM-\textit{Newton} will also play a role in the seasonal variation.   Some of these effects will be discussed in Section \ref{sec:seasonal_variation}.
 
 We can use our analytical predictions to explore the ($\kappa$, $T_0$) parameter range that is consistent with the observations shown in Fig.~\ref{fig:XMMFig14a}.
 For this purpose, we calculate the maximum likelihood $\mathcal{L}(\kappa, T_0)$ defined as:
\begin{equation}
\label{L}
    \mathcal{L}(\kappa, T_0)=\exp \left[-\frac{1}{2} \left(\frac{d-f(\kappa, T_0)}{\sigma}\right)^2 \right],
\end{equation}
where the data, $d$, and the model, $f(\kappa, T_0)$, are estimated at one particular frequency, $\omega$. We choose $\omega=3$~keV, but any frequency would work, since the model and the observations show a very similar frequency dependence.  The value of  $d$ in Eq. (\ref{L}) is chosen as the middle of the four observed spectra, which is defined as the average of the top spectrum (Fall) and bottom spectrum (Winter), i.e. $d\equiv \frac{1}{2}[f^{(\rm obs, F)}+f^{(\rm obs, W)}]$. The variance, $\sigma$, is chosen as $\sigma \equiv \frac{1}{2}[f^{(\rm obs, F)}-f^{(\rm obs, W)}]$, which represents the maximum signal variation between the four seasons. Note that we are not in a position to calculate a full likelihood function over all frequencies, since we do not know the correlation for different $\omega$, and the resulting likelihood would be difficult to interpret. Nevertheless, our approach should provide a reasonable order of magnitude estimate of the region of the parameter space, ($\kappa, T_0$), consistent with the observations. We are not trying to interpret $\mathcal{L}(\kappa, T_0)$ in a probabilistic way because our error estimate is only approximate. However, the maximum of $\mathcal{L}(\kappa, T_0)$ at $1$ is still a valid indicator of where the ($\kappa, T_0$) degeneracies lie. Fig.~\ref{fig:contour} shows the iso-contours of $2\ln\mathcal{L}(\kappa, T_0)$, where a lighter colour represents a better match.
%
%
The allowed parameter space is represented by two branches in Fig.~\ref{fig:contour}. The right vertical branch is essentially independent of $T_0$, and it matches the observations for $\kappa\sim 10^{-2.5}$. This ``insensitivity" to $T_0$ is consistent with the red and black lines in Fig. \ref{fig:Tt}, which illustrates the fact that AQN cooling is independent of $T_0$ when $\kappa$ is high enough. On the other hand, the left branch is strongly dependent on both $\kappa$ and $T_0$.
The next step is to investigate the seasonal variation in the context of our model. From the qualitative arguments given in Section \ref{subsec:cooling}, the physically preferred values for $\kappa$ and $T_0$ are in the right branch. However, for completeness, we will also calculate the seasonal variations for a lower value of $\kappa$. In the next section, the calculations will be restricted to the three sets of parameter values represented by the big solid dots in Fig.~\ref{fig:contour}.

  We conclude this section with the following comment.
Our computation of the spectrum (including the extension to much higher energies)  is robust and can be used to plan future experiments to perform the annual modulation studies in the near-Earth environment without any modulation due to the seasonal variations. 
In the next Section we discuss the impact of the Earth position (subsection \ref{Earth-position}) and the possible complications due to the telescope altitude and orientation (subsection \ref{subsec2:seasonal_variation}).

\section{Seasonal variation}\label{sec:seasonal_variation}
Up to this section, our focus was on the calculation of the average intensity of the AQN radiation spectrum, ignoring the seasonal variations. However, the seasonal variation was the most important  feature discovered by \cite{Fraser:2014wja} as the conventional paradigm assumes that there should be no any seasonal variations as mentioned  in Introduction.    The authors claimed an 11$\sigma$ confidence level detection of the seasonal variation in the 2-6 keV energy band, after removing all possible instrumental contamination and known astrophysical sources. They argued that known conventional astrophysical sources had been ruled out as a possible explanation of their signal. The main goal of this section is to explain how the seasonal variation might occur
in the AQN framework. We will find that an annual amplitude modulation on the order of 20-25$\%$ is expected. Interestingly, with conventional dark matter models (e.g. WIMPs), any seasonal variation is expected at a much lower level, on the order of 1-10$\%$ for most of the exprements carried out on Earth's surface when the effect is entirely due to the flux differences,  see, e.g., \cite{Freese:1987wu} and \cite{Freese:2013}.

\subsection{Effect of the Earth's position on its orbit}\label{Earth-position}
\begin{figure}
    \centering
    \includegraphics[width=0.8\linewidth]{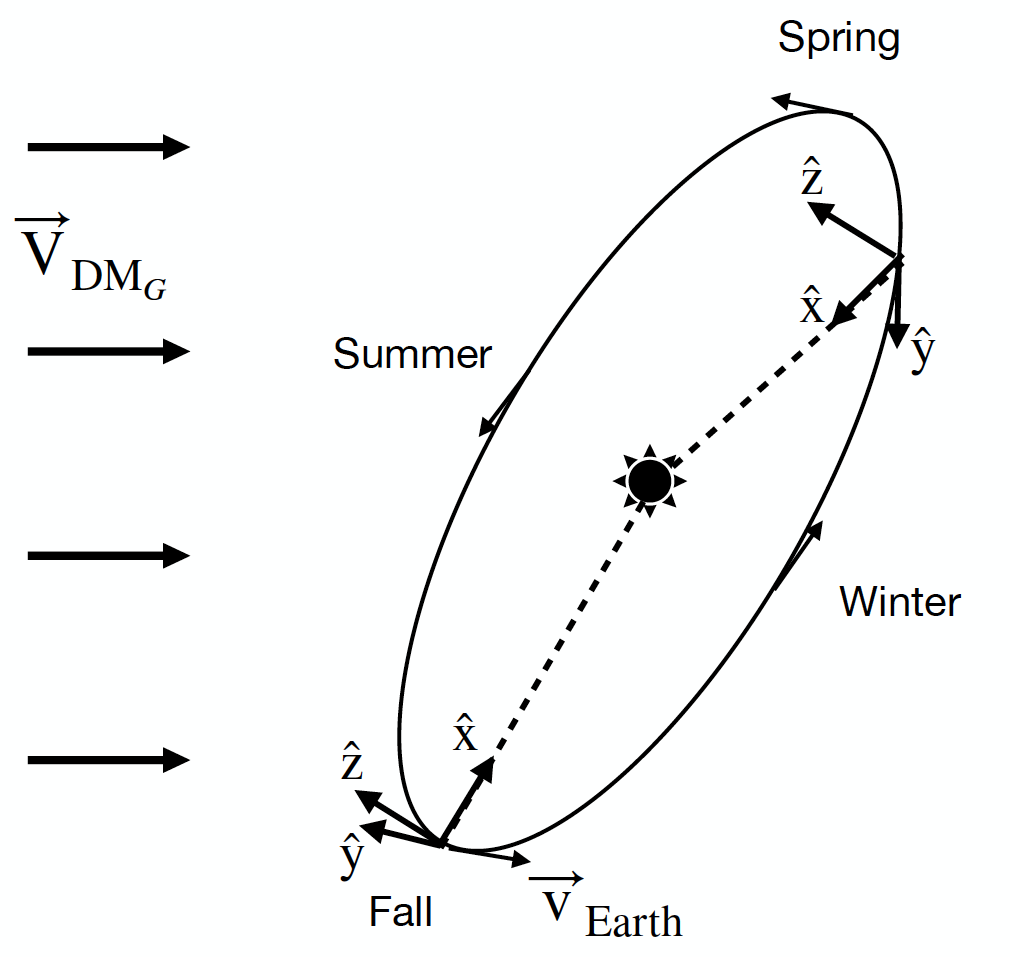}
    \caption{Motion of dark matter relative to the Solar System, which is taken as the fixed reference frame. The Earth moves in a nearly circular orbit, with a velocity, $\vec v_{\rm E}$, relative to the Sun. The location of the seasons relative to the orientation of the ecliptic plane and dark matter wind, $v_{{\rm DM}_{\rm G}}$, is important for the effect discussed in Section \ref{sec:seasonal_variation}.}
    \label{fig:solar_system}
\end{figure}

In Section \ref{sec:theoretical_predictions}, we introduced $v_{\rm in}$, the speed of a nugget hitting the Earth. However, the Earth's motion around the Sun leads to a seasonal variation of $v_{\rm in}$, which will affect the AQN signal given by Eq. (\ref{eq:dFrdomega}). In this subsection, we will calculate the amplitude of the seasonal effect, using the analytical prediction, Eq. (\ref{eq:dFrdomega}), and a realistic model of a nugget's incoming speed, $v_{\rm in}$, which is different in the Winter and Summer. 

The Sun is moving in the galactic plane, on a nearly circular orbit with velocity, $v_{{\rm DM}_{\rm G}}$, about the galactic center. The rotation of the dark matter halo is negligible compared to the rotation of the Sun. Therefore, the entire solar system is facing a dark matter wind with an average velocity of approximately $v_{{\rm DM}_{\rm G}}$. The tilt of the ecliptic plane relative to the dark matter wind is approximately $60 \degree$. This configuration is shown in Figure \ref{fig:solar_system}, along with the positions of the four seasons on the Earth's orbit.

The dark matter velocity with respect to the Sun is $\vec{v}_{\rm{DM}_{G}}$, with $v_{\rm{DM}_{G}} \simeq 220$~km/s. The velocity of the Earth around the Sun is $\vec{v}_{\rm E}$, with $v_{E}=30$~km/s. Consequently, the dark matter velocity with respect to the Earth is given by
\begin{equation}
    \vec{v}_{\rm in}=\vec{v}_{\rm{DM}_{G}}-\vec{v}_{\rm E},
\end{equation}
and the magnitude is
\begin{equation}\label{eq:v_magnitude}
    v_{\rm in}=\sqrt{v_{\rm{DM}_{G}}^2+v_{\rm E}^2-2v_{\rm{DM}_{G}}\cdot v_{\rm E}\cdot \cos{\theta}},
\end{equation}
where $\theta$ is the angle between $\vec{v}_{\rm E}$ and $\vec{v}_{\rm{DM}_{G}}$. $\theta$ is $60 \degree$ in the Winter, $120 \degree$ in the Summer, and $90 \degree$ in the Spring and Fall. Noting that $\vec{v}_{\rm{DM}_{G}}\gg \vec{v}_{\rm E}$, (\ref{eq:v_magnitude}) can be simplified via Taylor expansion as
\begin{equation}
     v_{\rm in}\simeq v_{\rm{DM}_{G}}-v_{E}\cos{\theta}.
\end{equation}
Using this approximation, the magnitude of the dark matter velocities (with respect to the Earth) in the four seasons is given by:
\begin{equation}\label{eq:v_seasons}
    \begin{aligned}
    &v_{\rm in}^{\rm (Sp)}=v_{\rm in}^{\rm (F)}\simeq v_{\rm{DM}_{G}},\\
    &v_{\rm in}^{\rm (W)}\simeq v_{\rm{DM}_{G}}-\Delta v,\\
    &v_{\rm in}^{\rm (S)}\simeq v_{\rm{DM}_{G}}+\Delta v,
    \end{aligned}
\end{equation}
where $\Delta v= \cos(60 \degree)\cdot v_{E}=15$~km/s. $\Delta v$ is the deviation from $220$~km/s, caused by the Earth's revolution around the Sun. The result is that the dark matter velocity is different for different seasons, as is apparent in Eq. (\ref{eq:v_seasons}), which leads to the seasonal variation of the X-ray background.

There are two specific features which are not shared by conventional WIMP models. The first one is related to the fact that the $v_{\rm in}$  and $v_{\rm out}$ velocities are different in the AQN model, but not in conventional dark matter. The second one is related to the fact that the intensity of the radiation explicitly depends on the number of nuggets which can be seen by the detector at each given moment, as shown in Fig. \ref{fig:xmmsketch} and computed in Eq. (\ref{eq:dFrdomega}). As we will see below, this leads to a feature unique to the AQN framework that is not shared by conventional dark matter.

The first effect, related to $v_{\rm out}\neq v_{\rm in}$, can be explained as follows. The passage of the AQN through the Earth is accompanied by friction and annihilation events with the surroundings, leading to $v_{\rm out} < v_{\rm in}$. We used $v_{\rm out}=v_{\rm in}=220$~km/s in Section~\ref{sec:theoretical_predictions}, a simplification that was sufficient to estimate the average of the AQN-induced spectrum. However, in this section, the fact that $v_{\rm out}$ is smaller than $v_{\rm in}$ may have an important impact on the seasonal variation. The reason is that as $v_{\rm out}$ gets closer to $\Delta v$, the seasonal variation becomes relatively more important. 
Different nuggets have different paths through the Earth, which results in different $v_{\rm out}$ even for the same $v_{\rm in}$. The precise distribution of $v_{\rm out}$ can only be obtained by numerical simulations, which is well beyond of the scope of the present work.
The speed change from crossing the Earth is quantified by the parameter $\gamma$:
\begin{equation}\label{eq:ratio_gamma}
    \gamma=\frac{v_{\rm out}}{v_{\rm in}},
\end{equation}
which is assumed to be constant in the present work, but may vary for different trajectories and different nugget's sizes.
Combining Eq. (\ref{eq:v_seasons}) with Eq. (\ref{eq:ratio_gamma}), we obtain the following expressions for $v_{\rm out}$ in the four seasons:
\begin{equation}\label{eq:vout_seasons}
    \begin{aligned}
    &v_{\rm out}^{\rm (Sp)}=v_{\rm out}^{\rm (F)}\simeq \gamma v_{\rm{DM}_{G}},\\
    &v_{\rm out}^{\rm (W)}\simeq \gamma(v_{\rm{DM}_{G}}-\Delta v),\\
    &v_{\rm out}^{\rm (S)}\simeq \gamma(v_{\rm{DM}_{G}}+\Delta v).
    \end{aligned}
\end{equation}

\exclude{
Our next task is to study  the dependence of the photon's spectrum  on $v_{\rm out}$. As we see from Fig. \ref{fig:counts}, there is a turning point on the AQN-induced spectrum. This is because the spectrum emitted by a nugget is a piecewise function (see Eq.~(\ref{eq:dFdomegaexpression})). Only the second branch (i.e., $\omega>\omega_{p}(z=0)$) of the piecewise function is important for us, because it decreases following a power law which matches the shape of the observed spectra. As we have explained in footnote~\ref{foot:cusp}, the cusp behavior in the left end of the AQN-induced spectrum has no physical significance but a reflection of our simplified treatment of the regions of small $\omega$. Thus, the analysis of seasonal variance in this section focuses on the second branch of the spectrum.
}

The second effect is due to the fact that the number of nuggets passing through the detection cone of the XMM-\textit{Newton} detector depends on $v_{\rm out}$ as well, as shown in Fig. \ref{fig:xmmsketch}. Using Eq. (\ref{eq:dFrdomega}), the \textit{average} $dF_{r}/d\omega$ measured by the detector is given by:
\begin{equation}\label{eq:dF_r_simplify}
    \frac{dF_{r}}{d\omega}\simeq n_{\rm AQN}(\bar{s})V \cdot \frac{dF}{d\omega}[T(\bar{s})],
\end{equation}
where $\bar{s}$ denotes the \textit{average} distance of the nuggets inside the cone, as viewed from the Earth's surface. The quantity $n_{\rm AQN}(\bar{s})$ is the number density of nuggets at distance, $\bar{s}$, while  $\frac{dF}{d\omega}[T(\bar{s})]$ is the spectrum emitted by a single nugget at distance, $\bar{s}$, determined by the temperature, $T(\bar{s})$. The volume, $V$, is the effective volume of the cone, which is a constant, inside which the nuggets contribute to the total spectrum received by XMM-\textit{Newton}. This means that we do not consider nuggets that are too far away from the detector. The detailed calculation of Eq. (\ref{eq:dF_r_simplify}) is shown in Appendix \ref{appendix:dFdw}, where we obtain an expression of $dF_{r}/d\omega$ as a function of $v_{\rm out}$:
\begin{equation}
\label{eq:dF_r_v_final_maintext}
    \frac{dF_{r}}{d\omega} \propto  \left(\frac{K_{1}}{v_{\rm out}} +K_{2}\right)^{-\frac{3.22}{c_{2}(\kappa)+3}},
\end{equation}
where $K_{1}$ and $K_2$ are functions of $\kappa$ and $T_{0}$ (see Appendix \ref{appendix:dFdw} for details).

The maximum seasonal difference is expected between Summer and Winter, because they have the maximum velocity difference, $2\gamma\Delta v$, as seen from Eq. (\ref{eq:vout_seasons}). We define the ratio
\be
\label{r}
r\equiv\frac{\left(\frac{dF_{r}}{d\omega}^{(\rm S)}\right)}{\left(\frac{dF_{r}}{d\omega}^{(\rm W)}\right)}
\ee
as the difference between the Summer and Winter spectra. Using Eq. (\ref{eq:vout_seasons}) and  Eq. (\ref{eq:dF_r_v_final_maintext}), we get
\begin{equation}\label{eq:ratio_SW}
    \begin{aligned}
        r\simeq  \left[\frac{\gamma^{-1}K_{1}/(v_{\rm{DM}_{G}}+\Delta v) +K_{2}}
    {\gamma^{-1}K_{1}/(v_{\rm{DM}_{G}}-\Delta v) +K_{2}}\right]^{-\frac{3.22}{c_{2}(\kappa)+3}},
    \end{aligned}
\end{equation}
where $v_{\rm{DM}_{G}}=220$~km/s and $\Delta v=15$~km/s, as discussed above. 

The functions $K_1$ and $K_2$ play a very important role in our study. If the temperature, $T$, strongly deviates from its initial value, $T_0$, such that $T\ll T_0$ at the moment of observation, then the second term with $T_0$ in the brackets in Eq. (\ref{T_0}) 
can be ignored, which drastically simplifies all equations.
In particular, the term $K_2\sim T_0^{-[c_{2}(\kappa)+3]}$ in Eq. (\ref{eq:K1K2})  can be neglected. This implies that $K_2$ in Eq. (\ref{eq:dF_r_v_final_maintext}) can be also ignored, which drastically simplifies the analysis.

\begin{figure}
    \centering
    \includegraphics[width=1\linewidth]{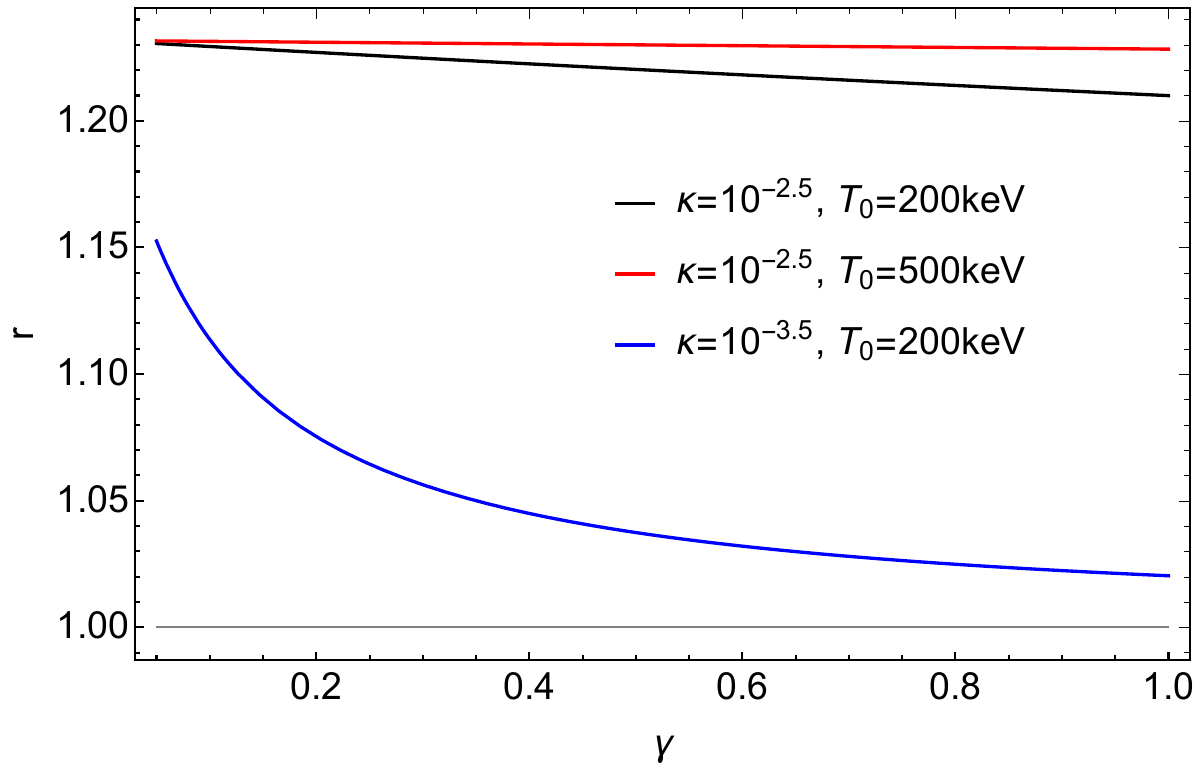}
    \caption{The ratio $r$ as a function of $\gamma$ for different groups of $\kappa$ and $T_0$. The solutions from the right branch from Fig. \ref{fig:contour} (red and black lines) will always produce $r\approx 20\%$ irrespectively to the value of $\gamma$, while the solution from left branch from Fig. \ref{fig:contour} (blue line) will   always generate a very small value of $r$. } 
    \label{fig:r_gamma}
\end{figure}

In this case, Eq. (\ref{eq:ratio_SW}) can be simplified to
\begin{equation}
\label{eq:ratio_SW_kappa1}
    r \simeq \left(\frac{v_{\rm{DM}_{G}}+\Delta v}{v_{\rm{DM}_{G}}-\Delta v}\right)^{\frac{3.22}{c_{2}(\kappa)+3}}\approx 1.23,
\end{equation}
which does not depend on $\gamma$.
This is a very solid and robust consequence of the AQN model. One should also emphasize that the condition $T\ll T_0$ is always  satisfied  for all solutions on the 
right branch shown in Fig \ref{fig:contour}. Indeed, the temperature, $T$, drastically drops for any value of $T_0$ with   $\kappa\simeq 10^{-2.5}$, as shown in Fig.~\ref{fig:Tt}. 

Eq. (\ref{eq:ratio_SW_kappa1}) is a very important result. It shows that for solutions from the right branch of Fig. \ref{fig:contour}, the seasonal variation could be large, up to $\sim$ 20-25$\%$, relatively insensitive to the exact value
of $\gamma$. Fig.~\ref{fig:r_gamma} shows the results of the exact computation from Eq. (\ref{eq:ratio_SW}) supporting this  claim, where the red and black lines remain relatively flat at $r\approx 1.23$ for all values of $\gamma$. The solutions from the left branch lead to a considerably smaller amplitude of the seasonal variation for any values of $\gamma$, as illustrated by the blue line in Fig.~\ref{fig:r_gamma}.
In the context of the AQN framework, the result (\ref{eq:ratio_SW_kappa1}) provides a strong argument in favour of a solution in the right branch of Fig. \ref{fig:contour}, because only the right one is capable of leading to seasonal variations in agreement with \cite{Fraser:2014wja}.

\exclude{
To see exactly the seasonal variance $r$, we plot the full expression (\ref{eq:ratio_SW}) in Fig.~\ref{fig:r_gamma} for different values of $\kappa$ and $T_0$\footnote{In plotting Fig.~\ref{fig:r_gamma}, the average distance $\bar{s}$ is chosen to be $6R_{E}$ to match the configuration Fig.~\ref{fig:xmmsketch}.}. We see that the $23\%$ seasonal variance, (\ref{eq:ratio_SW_kappa1}), is saturated for $\kappa=10^{-2.5}$.

Fig.~\ref{fig:r_gamma} clearly shows the seasonal variance caused by the Earth's revolution around the Sun within the framework of AQN model, but it is only for the case that all nuggets leave the Earth with the \textit{uniform} velocity ($\gamma$ fixed at a single value between $0.1$ and $1$, the same value for all nuggets). In reality, different nuggets have different out velocities averaged as $\left<\gamma\right>=1/2$, so the value of $r$ at $\gamma=1/2$ could better represent the seasonal variance for a given group of $\kappa$ and $T_{0}$.
}

\begin{figure}
    \centering
    \includegraphics[width=1\linewidth]{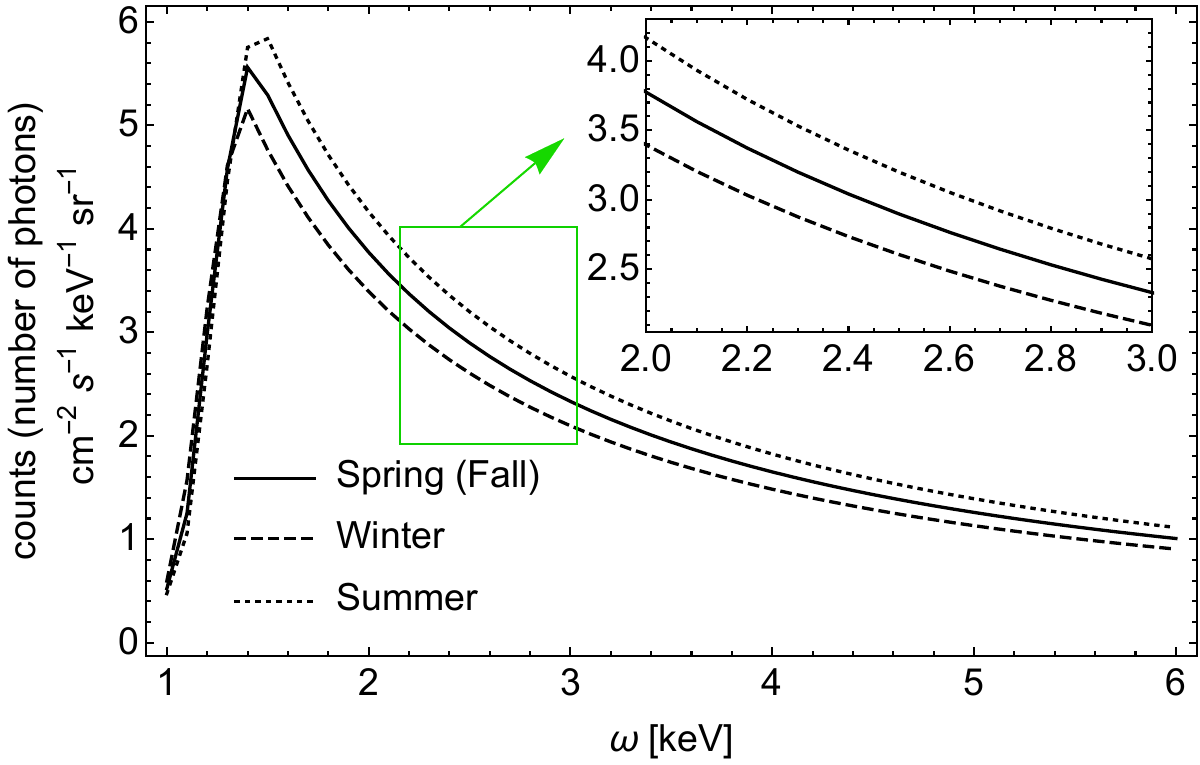}
    \caption{Demonstration of the seasonal variation   with specific parameters  $(\kappa,T_0, \gamma)=(10^{-2.5}, 200{\rm~keV}, 0.5)$ as an example. A small portion of the spectrum, $\omega\in$ 2-3 keV, is zoomed  in 
to demonstrate a large seasonal variation on the level of $\approx 20\%$.}
    \label{fig:f_gamma}
\end{figure}

As an example of seasonal variation, Fig.~\ref{fig:f_gamma} shows $f^{\rm (theory)}$, as defined  by Eq. (\ref{eq:f_xmm}), with $\frac{dF_{r}}{d\omega}$ given by Eq. (\ref{eq:dFrdomega}). For this plot, we choose $\kappa=10^{-2.5}, T_{0}=200$~keV,
and $\gamma=1/2$. However, as we have shown in the previous sections, the radiation spectrum is not very sensitive to parameters $T_0$ and $\gamma$, as long as we choose a solution from the right branch of Fig. \ref{fig:contour}.
A sample of the spectrum with $\omega\in$ 2-3 keV shows a large seasonal variation at the level of $\sim$ 20-25$\%$.

The maximal seasonal variation observed by  \cite{Fraser:2014wja} can be estimated from normalization factors, $N_0$, given by Eq. (\ref{N_0}) for different seasons as follows:
\be\label{N_obs}
\frac{N_0({\rm Fall})-N_0({\rm Winter})}{N_0({\rm Fall})+N_0({\rm Winter)}}\approx 0.29,
\ee
which is a very large effect. One should emphasize that the seasonal variation (\ref{N_obs}) cannot be directly compared with our estimate of parameter $r$, computed for Summer-Winter modulation (\ref{eq:ratio_SW_kappa1}). This is due to the satellite's positions and the orientations of the detector, which will be discussed in Subsection \ref{subsec2:seasonal_variation}. The main lesson of our computations is that the annual modulation effect is very large, much larger than conventional WIMP models can predict \citep{Freese:1987wu,Freese:2013}, which are on the level of 1-10$\%$ computed at the Earth's surface.

\exclude{
To see more intuitively the seasonal variance, in  we plot $f^{\rm (theory)}$ (Eq. (\ref{eq:f_xmm})) for four seasons where the velocities are given by (\ref{eq:vout_seasons}) with $\gamma$ fixed at the average value $\gamma=\left<\gamma\right>=1/2$. Fig.~\ref{fig:f_gamma} is plotted with $\kappa=10^{-2.5}, T_{0}=200$~keV as an example. It can also be used to test that our simplifications leading to eqs.  (\ref{eq:dF_r_v_final_maintext}), (\ref{eq:ratio_SW}) and (\ref{eq:ratio_SW_kappa1}) are correct by comparing it with Fig.~\ref{fig:r_gamma}.


Although Fig.~\ref{fig:r_gamma} and Fig.~\ref{fig:f_gamma} are limited to the simplified case that $\gamma$ is the same for all nuggets, they meet our goal in this section to analytically show the seasonal variance rather than give precise result which can only be completed by numerical simulations. We do the simulations in section~\ref{sec:simulations} which incorporate interactions between nuggets and the surroundings inside the Earth. Thus, the realistic pattern of $v_{\rm out}$ distribution as Fig.~\ref{fig:vout} will be automatically included there without any simplifications made in this section.
}

\subsection{Effect of the satellite's position and orientation on its orbit}\label{subsec2:seasonal_variation}

The previous calculations show that we should expect a seasonal modulation of the signal, which should be strongest in Summer, weakest in Winter, and equally half-way for Fall and Spring. This effect is entirely driven by the strength of the local dark matter wind speed. Compared to Fig. \ref{fig:XMMFig14a}, one can see that this does not quite agree with the seasonal modulation measured by \cite{Fraser:2014wja}, as given by (\ref{N_0}). In their measurement, the Fall amplitude is the highest, and Summer and Spring are equal. However, as noted by \cite{Fraser:2014wja}, there are two additional factors which can change the seasonal modulation of the X-ray background significantly: the altitude of the telescope and the direction of the field-of-view (or beam). These factors are particularly important in the context of our model. Firstly, the altitude plays a role because, as shown by Eq. (\ref{eq:dF_r_simplify}), a nugget's temperature upon exiting Earth decreases quickly with altitude. Therefore, if the telescope observes from a high altitude out, the X-ray background should be lower than if observed from a lower altitude. Secondly, the orientation of the telescope beam is also important. The X-ray background will indeed be stronger in the direction opposite to the incoming dark matter wind. This is caused by the fact that, on average, more heated nuggets will emerge from the side of the Earth opposite to the side where they preferentially entered from. 
Consequently, depending on the telescope position and orientation, the seasonal variation of the X-ray background can be altered. 
\exclude{ Fortunately, these are effects which can be completely accounted for, as long as the telescope's orbital parameters are known exactly. This is what \cite{Fraser:2014wja} have done in their study for their specific solar-axion model. While their model has a number of fundamental major problems, as mentioned in the introduction, the main point is that the seasonal variations as observed by XMM-\textit{Newton}  do not follow the standard annual modulation with a simple $\cos (\Omega t+\phi_0)$ form, as is normally expected \citep{Freese:1987wu, Freese:2013}.   In particular,  Figure 5  from \cite{Fraser:2014wja} shows that the seasonal variation predictions differ for different observing epochs, because of the particular telescope positions and orientations at those epochs. The resulting effect can significantly change the phase of the seasonal variation and the amplitude by approximately a factor of two.
}

The reason that these effects could be potentially  very strong is because the XMM-\textit{Newton} is placed on a Highly-elliptical Earth Orbit (HEO), with an inclination of $40\degree$ relative to the ecliptic plane, a southern apogee altitude of $\sim 115000$ km, and a perigee of $\sim 6000$ km, with an orbital period of $48$ hours. 
The exact prediction of the X-ray background becomes a highly non-trivial task, which requires precise knowledge  of the telescope position and orientation for every data point being taken, which is beyond  the scope of the present work. 

\exclude{At the same time, the obtained spectrum represents a very solid and robust result, which is not sensitive to
the telescope's  position and orientation. Furthermore, a strong seasonal variation (difference between maximum and minimum intensity) represented by Eq. (\ref{eq:ratio_SW_kappa1}) is also a very solid and robust property of the AQN framework, not sensitive to any specific details of the model. 
The comparison of our prediction in X-ray to the signal measured by \cite{Fraser:2014wja}  strongly  constrains  the  parameter  $\kappa$, and very mildly constrains the initial temperature, $T_0\in$ 200-500 keV. This is because nuggets with very different temperatures behave in a very similar way after the long journey of $t\geq 10^2$s, where XMM-\textit{Newton} is operational at distances of $r\gtrsim 8 R_{\oplus}$, as one can see from Fig. \ref{fig:Tt}.  However, the most important  message here is that the intensity, spectrum, and magnitude of the seasonal variation on the level 
of 20-25$\%$, measured by \cite{Fraser:2014wja}, can be naturally accommodated within the AQN framework, as argued in the present work. 
}
\section{Conclusion}\label{sec:conclusion}
 The observation \cite{Fraser:2014wja} of the seasonal variations in the Cosmic X-ray Background residual is a highly unexpected phenomenon. We propose an explanation which avoids the issues of the model based on solar axions proposed by \cite{Fraser:2014wja}. Our suggestion is that dark matter is made of AQNs which occasionally cross the Earth, and when they emerge they emit X-rays that contribute to the near-Earth X-ray background. The main results of the present work can be summarized as follows:
 \bigskip
 
 1. Our estimations  for the expected signal  are presented in Fig.\ref{fig:f_gamma}. In the 2-6 keV energy band, it is consistent with the intensity and  spectral shape observed  by \cite{Fraser:2014wja};
 
 2. We computed the spectrum and the intensity of the AQNs emission where  XMM-\textit{Newton} is operational, i.e. $r\gtrsim  8R_{\oplus}$.  The corresponding results are presented in Fig. \ref{fig:counts} and  shown by the solid lines;

 3. The obtained results are consistent with the energy spectrum observed by \cite{Fraser:2014wja}, which are also shown in Fig. \ref{fig:counts} with the four dashed lines (for the four seasons). The important point is that the shape of the spectrum is not sensitive to any details of the model, and represents very solid and robust predictions of the entire AQN framework. We did not attempt to reproduce the seasonal variation shown on Fig.~\ref{fig:XMMFig14a}) which requires the precise position and orientation of the telescope and is left for future work. This was explained in subsection \ref{subsec2:seasonal_variation};
 
 4. This spectrum extends to much higher energies, up to 100 keV. This should be considered as a very robust prediction of the AQN framework.  It can be tested in future experiments by any instrument sensitive to energies above 6 keV, representing the XMM-\textit{Newton}  cutoff energy\footnote{cutoff energy for EPIC pn is $\sim 15$ keV. However, the authors of \cite{Fraser:2014wja} do not provide any details for
 seasonal variations for higher energies above 6 keV.}.
 
 5. We also computed the parameter, $r$, which represents the maximal range of seasonal variations. We found that $r\approx$ 20-25$\%$, from Eq. (\ref{eq:ratio_SW_kappa1}).
 This result is not very sensitive to our parameters within our simplified treatment of the problem as discussed in the text. 
 
 6. The parameter $r$ describing the  seasonal variation remains large for much higher energies, up to 100 keV, as mentioned in item 2. This prediction can be tested in future experiments by any instrument sensitive to energies well above 6 keV. 

\bigskip
In this paper, we did not explore all of the possible masses and sizes that nuggets can have. In contrast to the uniform size, $R$, and the uniform velocity, $v_{{\rm DM}_{\rm G}}$, of AQNs used in the present work, the more realistic case is that the AQN size follows a distribution based on percolation theory~\citep{Ge:2019voa}, and the AQN velocity follows a Gaussian distribution~\citep{Lawson:2019cvy}. In the future, it would be very desirable to take into consideration these two distributions using full scale Monte Carlo simulations. 

  Our computations of the spectrum and intensity can be used to plan future experiments which could perform the annual modulation studies in the near-Earth environment as long as they can be done without additional complications related to the changes of the altitude of the telescope and the direction of the field-of-view. 
On the other hand, the satellite position and orientation can strongly affect the seasonal modulation of the X-ray signal, as we have discussed in Subsection~\ref{subsec2:seasonal_variation}. Thus, we need to know the exact positions and orientations of XMM-\textit{Newton} when it made those observations. With the full orbit information implemented into the AQN framework, we may finally fully reproduce the observed pattern in Fig.~\ref{fig:XMMFig14a}, i.e. Fall$>$Spring $\approx$ Summer$>$ Winter. We leave this for our future work.

Perhaps, the most important aspect of this work is our prediction for near-Earth seasonal variations at higher energies. Indeed, the CXB observed by XMM-Newton was used here to constrain a few parameters of the model, which leads to a deterministic prediction of the signal at higher energies. Such a prediction could provide a decisive test of the AQN model. As shown in this paper, the radiation spectrum extends well beyond 6 keV, and we are in a position to make a prediction in the $\gamma$-ray range. In particular, NuSTAR (Nuclear Spectroscopic Telescope Array) 
is a NASA space based X ray telescope which operates in the range 3 to 79 keV. It would be very desirable\footnote{ We thank anonymous Referee for this suggestion.} if NuSTAR could analyse  the seasonal variation in extended energy range up to 79 keV. Essentially this work predicts the spectrum and estimates  the intensity in this energy range. Another, possible detector which is capable to test our predictions is the Gamma-ray Burst Monitor (GBM) instrument on the Fermi Telescope which has multi-year archival data of $\gamma$-ray background measurements in the near-Earth environment \citep{meegan2009fermi}. This would constitute the ideal data set to test our model because we are able to predict uniquely the X-ray background, as seen with XMM-\textit{Newton}, and the $\gamma$-ray background, as seen by GBM. According to the AQN model, the two backgrounds, separated by two orders of magnitude in frequency, should share very similar properties, once the instrumental and astrophysical sources are removed. This exciting project is left for our future work.   


\section*{Acknowledgments}
This work was supported by the National Science and Engineering Research Council of Canada.
 
\appendix  
 
\section{Calculations of $\lowercase{d}F/\lowercase{d}\omega$ and $F_{\rm \lowercase{tot}}$}\label{appendix:F_details}
First, we calculate the spectral surface emissivity (\ref{eq:dFdomega0}) with all of the extra effects discussed in Section~\ref{subsec:radiation} included. Only photons with an energy larger than the plasma frequency, $\omega_{p}(z)$, can propagate outside of the system. The largest plasma frequency, $\omega_{p}(z=0)$, occurs in the deepest region of the electrosphere, where the positron density is the largest. Therefore, photons with an energy, $\omega>\omega_{p}(z=0)$, created anywhere in the electrosphere, ($z\geq0$), can propagate outside of the system. For $\omega<\omega_{p}(z=0)$, there is a cutoff determined by (\ref{eq:omega}):
\begin{equation}\label{eq:z0}
    z_{0}(\omega)=\frac{1}{\omega} \sqrt{\kappa} \sqrt{\frac{2T}{m_{e}}}-\bar{z}.
\end{equation}
Photons with an energy, $\omega<\omega_{p}(z=0)$, can propagate outside of the system only if they are created in the regime, $z>z_{0}(\omega)$. Therefore, $dF/d\omega$ (\ref{eq:dFdomega0}) becomes a piecewise function with $\omega_{p}(z=0)$ as the turning point. 

We should also notice that when $\omega$ is small enough, the lower cutoff, $z_{0}(\omega)$ (\ref{eq:z0}), could be larger than the upper cutoff, $z_{1}$ (\ref{eq:z_1}), defined by the ionization effect. We can then get a critical frequency by equating $z_{0}=z_{1}$:
\begin{equation}\label{eq:omega_z0z1}
    \omega_{z_{0}=z_{1}}(T)=\sqrt{\kappa} \sqrt{\frac{2T}{m_{e}}}[z_{1}(T)+\bar{z}(T)]^{-1}.
\end{equation}
We see that $z_{0}<z_{1}$ for $\omega>\omega_{z_{0}=z_{1}}$, while $z_{0}>z_{1}$ for $\omega<\omega_{z_{0}=z_{1}}$. Only photons with $\omega>\omega_{z_{0}=z_{1}}$ can be generated. The low frequency photons with $\omega<\omega_{z_{0}=z_{1}}$ cannot be generated because the region of the electrosphere that could generate them is ionized (see Eq. (\ref{kappa})). Therefore, $dF/d\omega$ should be written as:
\begin{equation}\label{eq:dFdomega}
    \frac{dF}{d\omega}(\omega)=
    \begin{cases}
    \frac{1}{2}\int_{z_{0}(\omega)}^{z_{1}}dz~\frac{d \tilde{Q}}{d\omega}(\omega,z), & \text{if $\omega_{z_{0}=z_{1}}<\omega<\omega_{p}(z=0)$}; \\
    ~\\
   \frac{1}{2}\int_{0}^{z_{1}}dz~\frac{d \tilde{Q}}{d\omega}(\omega,z), & \text{if $\omega>\omega_{p}(z=0)$}.
  \end{cases}
\end{equation}
Integrating $d \tilde{Q}/d\omega$ (\ref{eq:dQdomega}) over $z$ gives:

\begin{equation}\label{eq:dQ}
\begin{aligned}
    \int dz \frac{d\tilde{Q}}{d\omega}(\omega,z)
    &=\int dz~n^{2}(z){\rm e}^{-\omega_{p}(z)/T} G(\omega)\\
    &=\kappa^2 \left(\frac{T}{2\pi\alpha}\right)^2  G(\omega)  \int dz~\frac{ {\rm e}^{-\sqrt{\kappa} \sqrt{\frac{2}{m_{e}T}}\frac{1}{z+\bar{z}}}   }{(z+\bar{z})^4}\\
    &=\kappa^2 \left(\frac{T}{2\pi\alpha}\right)^2  G(\omega)  H(z),
\end{aligned}
\end{equation}
with 
\begin{equation}\label{eq:H}
\begin{aligned}
    H(z)={\rm e}^{-\sqrt{\frac{2\kappa}{m_{e}T}}\frac{1}{z+\bar{z}}} &\left[\frac{1}{\sqrt{\frac{2\kappa}{m_{e}T}}}\frac{1}{(z+\bar{z})^2} +\frac{2}{\left(\sqrt{\frac{2\kappa}{m_{e}T}}\right)^2}\frac{1}{(z+\bar{z})}\right.\\
    &+\left. \frac{2}{\left(\sqrt{\frac{2\kappa}{m_{e}T}}\right)^3}\right].
\end{aligned}
\end{equation}
$G(\omega)$ in (\ref{eq:dQ}) is a function defined for convenience to collect the terms that do not depend on $z$:
\begin{equation}
    G(\omega)\equiv \frac{4\alpha}{15}\left(\frac{\alpha}{m_{e}}\right)^2 2\sqrt{\frac{2T}{m_{e}\pi}} \left(1+\frac{\omega}{T}\right){\rm e}^{-\omega/T} h\left(\frac{\omega}{T}\right).
\end{equation}
The expression for $h(x)$ is:
\begin{equation}
    h(x)=17+12\left[\ln2+\left(1+{\rm e}^{x}\int_{1}^{\infty}\frac{{\rm e}^{-xy}}{y} dy \right)(1+x)^{-1}\right],
\end{equation}
which is a function derived in~\cite{Forbes:2008uf} (we refer the readers to Appendix A2 of~\cite{Forbes:2008uf} for further details). Plugging~(\ref{eq:dQ}) into (\ref{eq:dFdomega}), we get:
\begin{equation}\label{eq:dFdomegaexpression}
    \frac{dF}{d\omega}(\omega)=
    \begin{cases}
    \frac{1}{2}\kappa^2 \left(\frac{T}{2\pi\alpha}\right)^2 \cdot G(\omega)\cdot\left[H(z_{1})-H(z_{0}(\omega))\right], & ~ \\
     ~~~~~~~~~~~~\text{if $\omega_{z_{0}=z_{1}}<\omega<\omega_{p}(z=0)$}; &~\\
    ~\\
   \frac{1}{2}\kappa^2 \left(\frac{T}{2\pi\alpha}\right)^2 \cdot G(\omega)\cdot\left[H(z_{1})-H(0)\right], & ~ \\
   ~~~~~~~~~~~~\text{if $\omega>\omega_{p}(z=0)$}. & ~
  \end{cases}
\end{equation}
We plot $dF/d\omega$ vs. $\omega$ in Fig.~\ref{fig:dFdomega} of the main text for $\kappa = 10^{-3}$ with $T=20$~keV and $50$~keV respectively, for illustrative purpose.

Now, we are ready to calculate the total surface emissivity, $F_{\rm tot}(T)$, by integrating $dF/d\omega$ over $\omega$:
\begin{equation}\label{eq:Ftot}
\begin{aligned}
    F_{\rm tot}(T)&=\int_{\omega_{z_{0}=z_{1}}(T)}^{\infty}d\omega~\frac{dF}{d\omega}(\omega) \\
    &=\left[\int_{\omega_{z_{0}=z_{1}}(T)}^{\omega_{p}(z=0)} d\omega~\frac{1}{2} \int_{z_{0}(\omega)}^{z_{1}}dz\frac{dQ}{d\omega}(\omega,z)\right] \\
    &~~~~~~~ +\left[\int_{\omega_{p}(z=0)}^{\infty} d\omega~\frac{1}{2} \int_{0}^{z_{1}}dz\frac{dQ}{d\omega}(\omega,z)\right] \\
    &=\frac{\alpha}{15\pi^{5/2}}\frac{T^{5}}{m_{e}} \kappa^2 \left[I_{1}(T)+I_{2}(T)\right],
\end{aligned}
\end{equation}
with
\begin{equation}
\begin{aligned}
    &I_{1}(T)=\frac{1}{T}\sqrt{2}(m_{e}T)^{-3/2} \\ 
    & \times
    \int_{\omega_{z_{0}=z_{1}}(T)}^{\omega_{p}(z=0)} d\omega~
    \left(1+\frac{\omega}{T} \right)
    {\rm e}^{-\frac{\omega}{T}}
    h \left(\frac{\omega}{T} \right)
    \cdot\left[H(z_{1})-H(z_{0}(\omega))\right],\\
    &I_{2}(T)=\frac{1}{T}\sqrt{2}(m_{e}T)^{-3/2} \\
    & \times
    \int_{\omega_{p}(z=0)}^{\infty} d\omega~
     \left(1+\frac{\omega}{T} \right)
     {\rm e}^{-\frac{\omega}{T}}
     h \left(\frac{\omega}{T} \right)
     \cdot\left[H(z_{1})-H(0)\right].
\end{aligned}
\end{equation}
The two dimensionless functions $I_{1}(T)$ and $I_{2}(T)$ can be solved numerically. 

\begin{figure}
    \centering
    \includegraphics[width=1\linewidth]{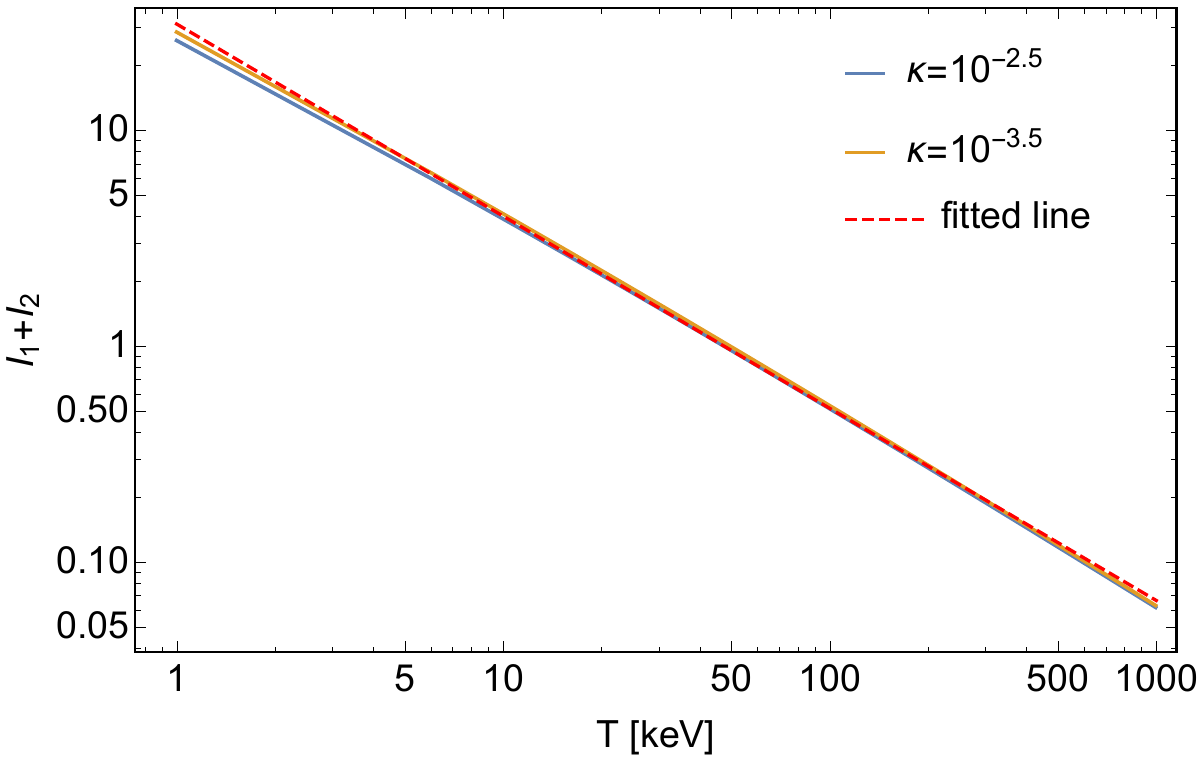}
    \caption{$[I_{1}(T)+I_{2}(T)]$ as a function of $T$, for $\kappa=10^{-2.5}, 10^{-3.5}$. We see that the two lines almost overlap with each other, and that they are fitted to the red dashed line.}
    \label{fig:I1I2}
\end{figure}

\begin{figure}
    \centering
    \includegraphics[width=1\linewidth]{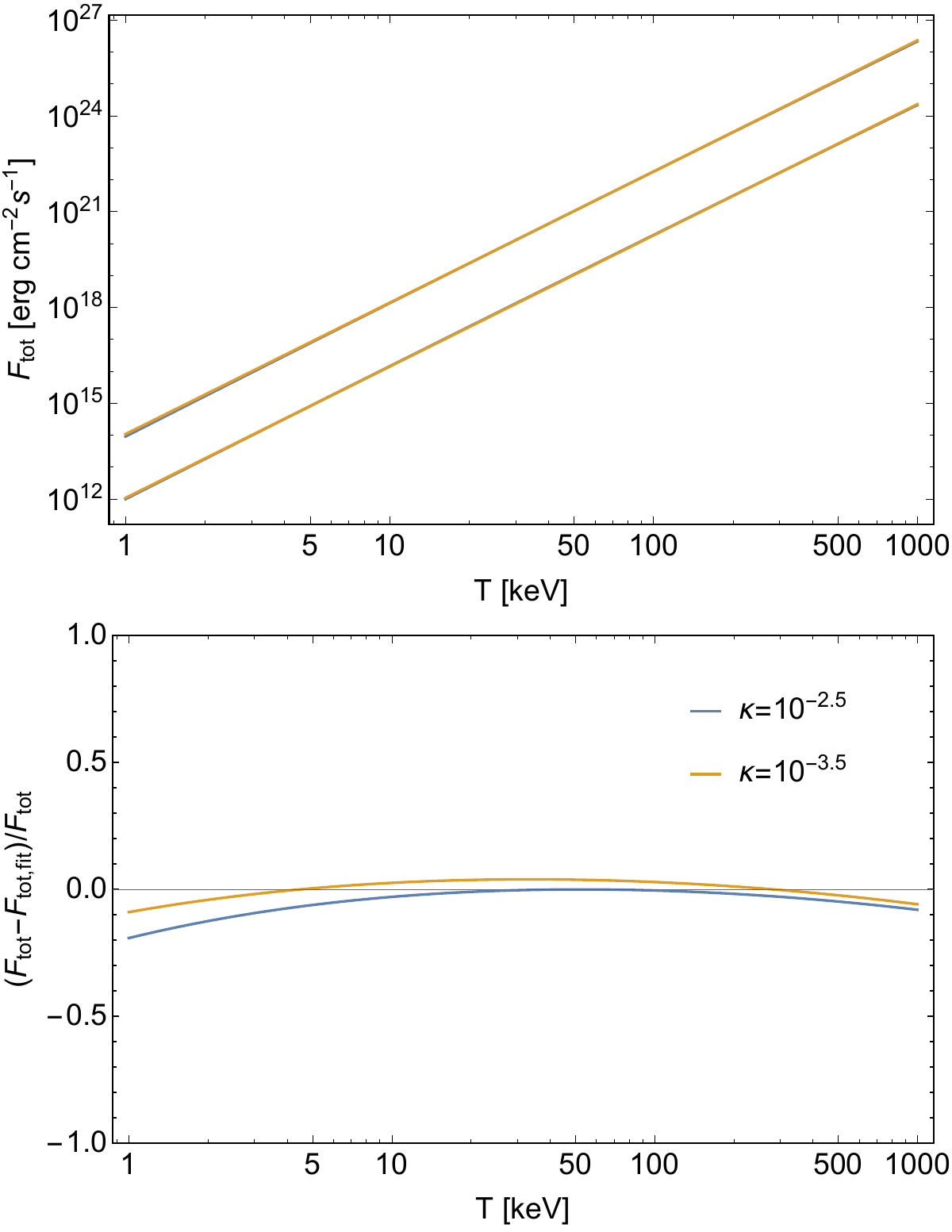}
    \caption{Top subfigure: the relation $F_{\rm tot}$ vs. $T$, for $\kappa=10^{-2.5}, 10^{-3.5}$ (top and bottom respectively). The blue lines are the exact $F_{\rm tot}$ (\ref{eq:Ftot}); the yellow lines are the fitted result (\ref{eq:Ftotfittingappendix}). We see that for each given $\kappa$, the blue line almost overlaps with the corresponding yellow line. Bottom subfigure: the relative error.}
    \label{fig:Ftoterror}
\end{figure}

In Fig.~\ref{fig:I1I2}, we plot $[I_{1}(T)+I_{2}(T)]$ vs. $T$ in the range $1~{\rm keV}\leq T\leq 1000~{\rm keV}$, for $\kappa=10^{-2.5}, 10^{-3.5}$ respectively. We see that the two lines of $[I_{1}(T)+I_{2}(T)]$, with $\kappa=10^{-2.5}$ and $10^{-3.5}$, almost overlap with each other, and that they are nearly a linear function of $T$ in the log-log scale. Then, we fit $[I_{1}(T)+I_{2}(T)]$ to a simple function (the red dashed line in Fig.~\ref{fig:I1I2}):
\begin{equation}\label{eq:I1I2fitting}
    [I_{1}(T)+I_{2}(T)]=c_{1}'\left(\frac{T}{10{\rm~keV}}\right)^{c_{2}'},
\end{equation}
with the two fitting parameters
\begin{equation}\label{eq:c12prime}
    c_{1}'=4,~~~~ c_{2}'=-0.89.
\end{equation}
This is a good approximation for $\kappa=10^{-2.5}, 10^{-3.5}$. Then, plugging (\ref{eq:I1I2fitting}) and (\ref{eq:c12prime}) into (\ref{eq:Ftot}), we get:
\begin{equation}\label{eq:Ftotfittingappendix}
    F_{\rm tot, fit}(T)=\frac{\alpha}{15\pi^{5/2}}\frac{T^{5}}{m_{e}} \kappa^2 \cdot c_{1}'\left(\frac{T}{10{\rm~keV}}\right)^{c_{2}'}.
\end{equation}
To see how good the fitted result (\ref{eq:Ftotfittingappendix}) is, we plot it together with the exact $F_{\rm tot}$ (\ref{eq:Ftot}) in the top subfigure of Fig.~\ref{fig:Ftoterror} for $\kappa=10^{-2.5}, 10^{-3.5}$. In the bottom subfigure of Fig.~\ref{fig:Ftoterror}, we also plot the relative error $(F_{\rm tot}-F_{\rm tot, fit})/F_{\rm tot}$. We see that the relative error is within $10\%$ for $T\gtrsim10$~keV.

\section{Calculations of nugget cooling}\label{appendix:cooling}

Solving the differential equation (\ref{eq:dTdt}) gives:
\begin{equation}
\begin{aligned}
\label{T_0}
\frac{t}{1{\rm~sec}}&\simeq \frac{R_{\rm AQN}}{1{\rm~sec}} \frac{5\pi^{5/2}}{3\alpha c_{1}(\kappa)[c_{2}(\kappa)+3]}
\frac{m_{e}(\mu_{u}^2+\mu_{d}^2)}{(10~{\rm keV})^{3}}\\
&~~~~~\cdot \left[\left(\frac{T}{10{\rm~keV}}\right)^{-[c_{2}(\kappa)+3]}
-\left(\frac{T_0}{10{\rm~keV}}\right)^{-[c_{2}(\kappa)+3]}\right]\\
&\simeq 
\frac{0.34}{c_{1}(\kappa)[c_{2}(\kappa)+3]}
\left(\frac{R_{\rm AQN}}{10^{-5}{\rm~cm}}\right) \left(\frac{\mu_{u,d}}{500{\rm~MeV}}\right)^2\\
&~~~~~\cdot \left[\left(\frac{T}{10{\rm~keV}}\right)^{-[c_{2}(\kappa)+3]}
-\left(\frac{T_0}{10{\rm~keV}}\right)^{-[c_{2}(\kappa)+3]}\right],
\end{aligned}
\end{equation}
or equivalently:
\begin{equation}\label{eq:T_t_appendix}
\begin{aligned}
    & T(t)\simeq  10{\rm ~keV}\cdot 
     \left[\frac{t}{1 {\rm~sec}}\left(\frac{R_{\rm AQN}}{10^{-5}{\rm~cm}}\right)^{-1}
    \left(\frac{\mu_{u,d}}{500{\rm~MeV}}\right)^{-2} \right.\\
    &\cdot \left(\frac{0.34}{c_{1}(\kappa) [c_{2}(\kappa)+3]}\right)^{-1} 
    \left. +\left(\frac{T_0}{10{\rm~keV}}\right)^{-[c_{2}(\kappa)+3]}\right]^{-\frac{1}{c_{2}(\kappa)+3}}.
\end{aligned}    
\end{equation}

\section{Calculations of $\lowercase{d}F_{\lowercase{r}}/\lowercase{d}\omega$ as a function of $\lowercase{v}_{\rm \lowercase{out}}$}\label{appendix:dFdw}


In this appendix, we are going to calculate (\ref{eq:dF_r_simplify}) to find the relation between $dF_{r}/d\omega$ and $v_{\rm out}$. First, we analyze the factor $dF/d\omega$ that occurs in (\ref{eq:dF_r_simplify}). The expression of $dF/d\omega$ is given in (\ref{eq:dFdomegaexpression}). As we have explained in Section~\ref{sec:seasonal_variation} of the main text, we are only interested in the second branch ($\omega>\omega_{p}(z=0)$) of the piecewise function (\ref{eq:dFdomegaexpression}). 

\begin{figure}
    \centering
    \includegraphics[width=1\linewidth]{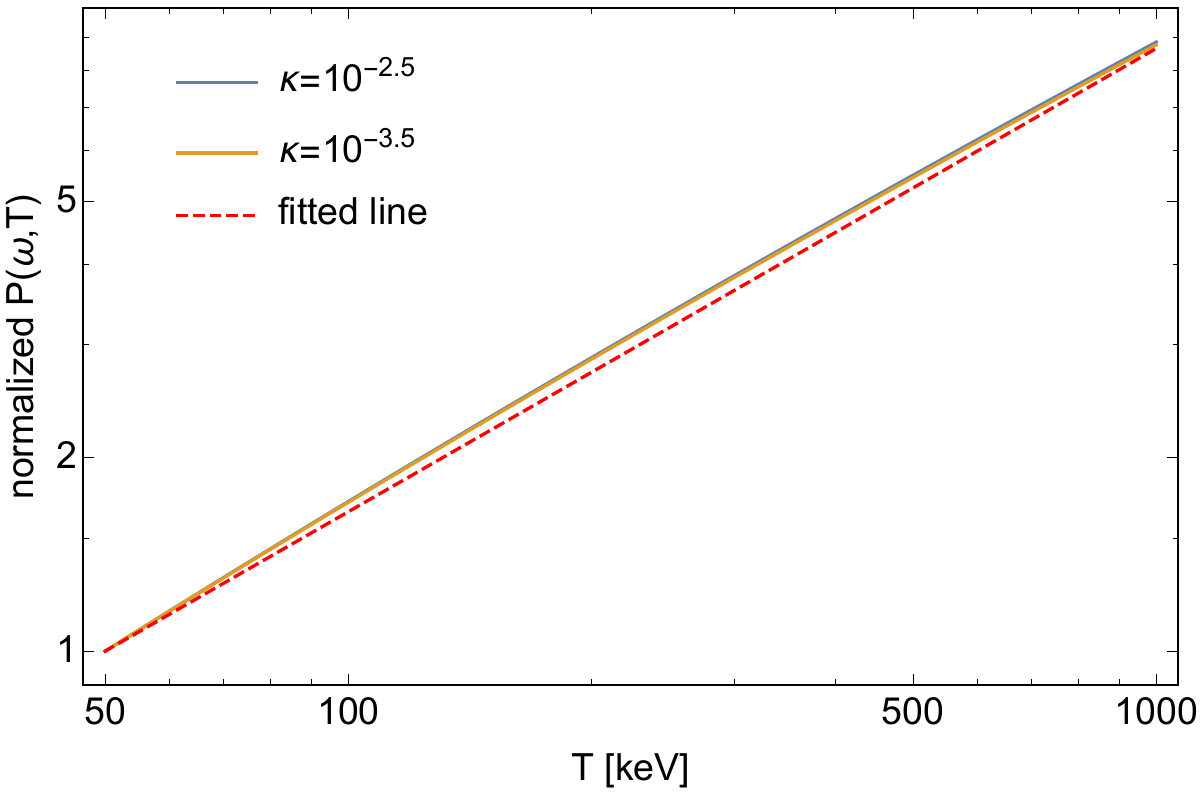}
    \caption{The relation between the normalized $P(\omega,T)$ and $T$, for $\kappa=10^{-2.5},10^{-3.5}$. The two lines almost overlap with each other. We use the function $[{\rm constant}\times T^{0.72}]$ (red dashed line) to fit the two lines. The two lines are plotted at $\omega=3$~keV. Changing the value of $\omega$ only slightly affects the relation between $P(\omega,T)$ and $T$. Since we do not care about the magnitude of $P(\omega,T)$, it is actually plotted in the normalized form, $P(\omega,T)/P(\omega,50{\rm~keV})$.}
    \label{fig:PT}
\end{figure}

As we can see from Fig.~\ref{fig:Tt}, the nuggets are still very hot when they enter the XMM-\textit{Newton}'s cone. We have $T\gg \omega$, where $\omega\sim$ 2-6 keV is the frequency range that we are interested in. This results in the pattern from the ``soft photon theorem,'' as explained in Section~\ref{sec:theoretical_predictions}. We can drop the terms suppressed by $\omega/T$, so the second branch of (\ref{eq:dFdomegaexpression}) is approximated as:
\begin{equation}\label{eq:dFdomega_P}
    \frac{dF}{d\omega}\propto T^{5/2} \cdot  P(\omega,T),
\end{equation}
where
\begin{equation}
    P(\omega, T)\equiv h\left(\frac{\omega}{T}\right) \left[H(z_{1}(T),T)-H(0,T)\right].
\end{equation}
In Fig.~\ref{fig:PT}, we plot the relation $P(\omega, T)$ vs. $T$, for $\kappa=10^{-2.5}, 10^{-3.5}$. We see that $P(\omega, T)$ can be well fitted to the red dashed line, which represents the function [${\rm constant}\times T^{0.72}$]. So we have:
\begin{equation}\label{eq:PT}
    P(\omega,T)\propto T^{0.72}.
\end{equation}
Note that $P(\omega,T)$ is also a function of $\omega$, which is only contained in $h(\omega/T)$. In plotting Fig.~\ref{fig:PT}, $\omega$ is chosen to be $3$~keV. Since $\omega/T\ll 1$, changing the value of $\omega$ only slightly affects the value of $P(\omega,T)$. Thus, to study the relation between $P(\omega, T)$ and $T$, we can fix $\omega$ at a certain value. This is good enough for our approximate analysis in this appendix. Combining (\ref{eq:dFdomega_P}) and (\ref{eq:PT}), we get:
\begin{equation}\label{eq:dF_T}
     \frac{dF}{d\omega}\propto T^{3.22}.
\end{equation}

The relation between $T$ and $v_{\rm out}$ is given in (\ref{eq:T_t_appendix}). We can rewrite (\ref{eq:T_t_appendix}) as:
\begin{equation}\label{eq:T_rewrite}
T\simeq  10{\rm ~keV}\cdot 
     \left[\frac{K_{1}(\kappa)}{v_{\rm out}} +K_{2}(\kappa,T_0)\right]^{-\frac{1}{c_{2}(\kappa)+3}},
\end{equation}
where
\begin{equation}\label{eq:K1K2}
\begin{aligned}
&K_{1}(\kappa)\equiv \frac{s}{1{\rm~sec}}\cdot \left(\frac{0.34}{c_{1}(\kappa) [c_{2}(\kappa)+3]}\right)^{-1}
\left(\frac{R_{\rm AQN}}{10^{-5}{\rm~cm}}\right)^{-1}\\
&~~~~~~~~~~~~~~~~\cdot
    \left(\frac{\mu_{u,d}}{500{\rm~MeV}}\right)^{-2},\\
&K_{2}(\kappa,T_0)\equiv \left(\frac{T_0}{10{\rm~keV}}\right)^{-[c_{2}(\kappa)+3]}.
\end{aligned}    
\end{equation}

Next, we check on possible variations of the factor, $n_{\rm AQN}$, that occurs in (\ref{eq:dF_r_simplify}). From ~(\ref{eq:nAQN}), we know that:
\begin{equation}\label{eq:n_v}
    n_{\rm AQN}\propto \frac{\mathcal{F}}{v_{\rm out}}\propto \frac{\mathcal{F}}{v_{\rm in}} \frac{v_{\rm in}}{v_{\rm out}} \propto \frac{v_{\rm in}}{v_{\rm out}}= \gamma^{-1},
\end{equation}
where we have used the relation that $\mathcal{F}/v_{\rm in}$ is a constant (see Eq. (\ref{F})), and  $\gamma$ is the ratio between $v_{\rm out}$ and $v_{\rm in}$, which is defined in Section~\ref{sec:seasonal_variation}. For simplicity, we assume that the loss of nugget velocity inside the Earth is proportional to the magnitude of the entry velocity, $v_{\rm in}$, so $\gamma$ and thus $n_{\rm AQN}$ are seasonally invariant, despite the fact that $v_{\rm in}$ changes with seasons.

Plugging (\ref{eq:dF_T}) and (\ref{eq:T_rewrite}) into (\ref{eq:dF_r_simplify}), we finally arrive at:
\begin{equation}\label{eq:dF_r_v_final}
    \frac{dF_{r}}{d\omega} \propto  \left(\frac{K_{1}}{v_{\rm out}} +K_{2}\right)^{-\frac{3.22}{c_{2}(\kappa)+3}}.
\end{equation}


\bibliographystyle{elsarticle-num.bst}

\bibliography{library}

\end{document}